\documentclass[11pt,a4paper]{article}

\usepackage{epsfig}
\usepackage{amsmath}
\usepackage{amssymb}
\usepackage{bbm}
\usepackage[usenames,dvipsnames]{xcolor}
\usepackage{verbatim}
\usepackage{bm}
\usepackage{color}
\usepackage{dsfont}
\usepackage{caption}
\usepackage[subrefformat=parens,labelformat=parens]{subfig}
\usepackage{subfloat}
\usepackage{cancel}
\usepackage{cite}
\numberwithin{equation}{section}

\newcommand{\beq}{\begin{equation}}
\newcommand{\eeq}{\end{equation}}

\newcommand{\bea}{\begin{eqnarray}}
\newcommand{\eea}{\end{eqnarray}}

\def\gsim{\mathrel{\raise.3ex\hbox{$>$\kern-.75em\lower1ex\hbox{$\sim$}}}}

\usepackage{booktabs}

\def\d2{W''}

\definecolor{lightcoral}{rgb}{0.94, 0.5, 0.5}
\definecolor{mahogany}{rgb}{0.75, 0.25, 0.0}
\definecolor{wildstrawberry}{rgb}{1.0, 0.26, 0.64}

\begin{document}

\begin{flushleft}
DESY 15-014\\
CPHT-RR002.0115\\
January 2015
\end{flushleft}

\vskip 1cm

\begin{center}
{\Large\bf Challenges for Large-Field Inflation and Moduli Stabilization} 

\vskip 2cm

{Wilfried Buchm\"uller$^a$, Emilian Dudas$^{b}$, Lucien Heurtier$^{b}$, Alexander Westphal$^a$,\\ Clemens Wieck$^a$}, Martin Wolfgang Winkler$^c$\\[3mm]
{\it{
$^a$ Deutsches Elektronen-Synchrotron DESY, 22607 Hamburg, Germany\\
$^b$ CPhT, Ecole Polytechnique, 91128 Palaiseau Cedex, France \\
$^c$ Bethe Center for Theoretical Physics and\\ Physikalisches Institut der Universit\"at Bonn, Nussallee 12, 53115 Bonn, Germany}}
\end{center}

\vskip 1cm

\begin{abstract}
\noindent 
We analyze the interplay between K\"ahler moduli stabilization and chaotic inflation in supergravity. While heavy moduli decouple from inflation in the supersymmetric limit, supersymmetry breaking generically introduces non-decoupling effects. These lead to inflation driven by a soft mass term, $m_\varphi^2 \sim m m_{3/2}$, where $m$ is a supersymmetric mass parameter. This scenario needs no stabilizer field, but the stability of moduli during inflation imposes a large supersymmetry breaking scale, $m_{3/2} \gg H$, and a careful choice of initial conditions. This is illustrated in three prominent examples of moduli stabilization: KKLT stabilization, K\"ahler Uplifting, and the Large Volume Scenario. Remarkably, all models have a universal effective inflaton potential which is flattened compared to quadratic inflation. Hence, they share universal predictions for the CMB observables, in particular a lower bound on the tensor-to-scalar ratio, $r \gtrsim 0.05$.
\end{abstract}

\thispagestyle{empty}

\newpage

\tableofcontents

%
\newpage
\section{Introduction}

Large-field chaotic inflation is an attractive scenario for describing the initial phase of the universe \cite{Linde:1983gd}. The simple quadratic potential $V = \frac12 m_\varphi^2\varphi^2$ predicts a scalar spectral index of $n_\text s \approx 0.967$ and a tensor-to-scalar ratio of $r \approx 0.13$ for 60 e-folds of inflation. Pure quadratic inflation is disfavored at the 2$\sigma$-level by observations of the Cosmic Microwave Background (CMB) \cite{Ade:2013uln,Ade:2015tva}. Here we embed quadratic inflation into supergravity with moduli stabilization. This leads to modifications such that quadratic inflation remains a viable possibility. The amplitude of scalar perturbations requires a small inflaton mass, $m_\varphi \approx 6 \times 10^{-6}$, and super-Planckian values of the inflaton field, $\varphi \lesssim 15$, during the slow-roll period.\footnote{Throughout this paper we work in units where $M_\text{P} = 1$.} At such large energy scales there are several good reasons to believe that the underlying theory should be space-time supersymmetric. However, the simplest implementations of chaotic inflation in supergravity are subject to a number of subtleties. 

In supergravity $\varphi$ is part of a complex scalar field $\phi = \frac{1}{\sqrt{2}} (\chi + i \varphi)$. The supergravity $\eta$-problem then requires the K\"ahler potential to have a symmetry to protect the inflaton from becoming too heavy. A simple candidate seems to be the global shift symmetry ${\phi \to \phi + i c}$, first used in chaotic inflation in \cite{Kawasaki:2000yn}. Clearly, the symmetry must be broken for the inflaton to be massive. Naively, a simple supergravity model can be defined by  
\begin{align}
 K = \frac{1}{2} \left(\phi + {\bar \phi}\right)^2 \,, \qquad
 W = \frac12 m \phi^2 \,. \label{i1}
\end{align}

There is, however, another problem. In this formulation the scalar potential is unbounded from below at large inflaton field values. Only for $\varphi \ll 1$ the potential defined by Eqs.~\eqref{i1} is approximately quadratic. This problem may be solved by invoking a second chiral multiplet, sometimes called ``stabilizer field'', denoted by $S$ \cite{Kawasaki:2000yn}. It is supposed to be heavier than the inflaton and to have a vanishing, or very small, vacuum expectation value during inflation. The simplest example is defined by 
\begin{align}\begin{split}
K &= \frac{1}{2}  (\phi + {\bar \phi})^2 + |S|^2 -  \frac{1}{\Lambda^2}|S|^4  + \dots\,, \\
W &= m S \phi \,. \label{ié}
\end{split}\end{align}
The quartic term in the K\"ahler potential is necessary for $S$ to be heavy enough during inflation. In particular, $m_S^2 \sim \frac{m^2 \varphi^2}{\Lambda^2} \sim\frac{H^2}{\Lambda^2}$, where $H$ denotes the Hubble scale during inflation. For $\Lambda \ll 1$ indeed $m_S \gg H$, while $\langle S \rangle = 0$. The inflaton potential is then simply
\begin{align}
V(\varphi) = \frac12 m^2 \varphi^2\,.
\end{align}

With such high energy scales involved it seems natural to study how a supergravity model of chaotic inflation can be embedded in string theory. There has been substantial progress in implementing chaotic inflation without a stabilizer, and related models, in string theory. For recent discussions, cf.~\cite{Palti:2014kza,Marchesano:2014mla,Hebecker:2014eua,Grimm:2014vva,Ibanez:2014kia,Blumenhagen:2014nba,Hebecker:2014kva,Ibanez:2014swa,Garcia-Etxebarria:2014wla}. In particular, the authors of \cite{Blumenhagen:2014nba,Hebecker:2014kva} have analyzed the effects of moduli stabilization in F-term axion monodromy inflation. A general supergravity analysis comparing the scale of inflation and the gravitino mass has been performed in \cite{Covi:2008ea,Covi:2008cn}. On the other hand, it has proven difficult to implement the model proposed in \cite{Kawasaki:2000yn} in explicit string constructions. For recent treatments, cf.\cite{Marchesano:2014mla,Dudas:2014pva}, and for a different approach, cf.~\cite{Mazumdar:2014bna}. Since string compactifications on Calabi-Yau (CY) manifolds typically yield an abundance of scalar fields in four dimensions, such as geometric moduli or the axio-dilaton, one may ask whether some of these fields can mitigate the problems of the quadratic inflation model without a stabilizer field. In particular, it may be possible that in no-scale supergravity setups involving moduli fields the negative term which makes $V$ unbounded from below is canceled. In the following we consider models which only contain K\"ahler moduli, assuming all other moduli to be stabilized supersymmetrically.

A no-scale cancellation, however, can only happen when the moduli break supersymmetry. In the absence of supersymmetry breaking, fields heavier than the Hubble scale can be completely decoupled from the dynamics of inflation, as discussed in \cite{Buchmuller:2014vda} and, for the case of chaotic inflation without a stabilizer field, in \cite{Buchmuller:2014pla}. On the contrary, supersymmetry breaking induces effects which do not decouple, in particular soft-breaking terms. Therefore, moduli stabilization with broken supersymmetry affects inflation even if the involved fields are heavy and can be integrated out. We divide moduli stabilization schemes in two classes.
\begin{enumerate}

\item The stabilization of moduli does not (or almost not) induce supersymmetry breaking. This means the moduli masses and the inflaton mass are much bigger than the scale of supersymmetry breaking, given by the gravitino mass $m_{3/2}$. In this case, the moduli can decouple with little effects on the dynamics of inflation, cf.~\cite{Buchmuller:2014vda}. Examples in this class are those with ``strong moduli stabilization'', treated in \cite{Kallosh:2004yh,Dudas:2012wi}, as well as stabilization via world-sheet instanton couplings as discussed in \cite{Wieck:2014xxa}. In models of this class chaotic inflation without a stabilizer does not work because the inflaton potential remains unbounded from below for $\varphi \gtrsim 1$. 

\item The stabilization of moduli spontaneously breaks supersymmetry such that the scale of supersymmetry breaking is larger than the inflaton mass. In this case, integrating out the heavy moduli results in substantial effects on the dynamics of inflation. This class is the main subject of this paper. As examples we study the model of KKLT \cite{Kachru:2003aw}, K\"ahler Uplifting \cite{Balasubramanian:2004uy,Westphal:2006tn}, the Large Volume Scenario \cite{Balasubramanian:2005zx}, and their interplay with chaotic inflation as defined in \eqref{i1}. In all three examples inflation is possible if the gravitino mass is larger than the Hubble scale. Many of the details are, however, different in the three cases. Note that the considered models of moduli stabilization are hardly compatible with the alternative inflation model of \cite{Kawasaki:2000yn} which requires the gravitino mass to be parametrically smaller than the inflaton mass \cite{Buchmuller:2014pla}.

\end{enumerate}

Despite the differences in detail, all considered models reduce to an effective single-field inflaton potential of remarkable universality. The moduli backreact on the inflaton, and the flattened effective potential in all models is of the form
\begin{align}
V=\frac12 m_\varphi^2 \, \varphi^2\,\left(1-\frac{\varphi^2}{2\varphi_\text M^2}\right)\,.
\end{align}
This potential is characterized by the inflaton mass $m$ setting the scale of the potential and position $\varphi_\text{M}$ of a local maximum induced by the negative quartic terms stemming from integrating out the moduli. Hence, all our setups share universal predictions for the CMB observables, in particular $r \gtrsim 0.05$, after imposing the Planck constraints.

This paper is organized as follows. In Sec.~2 we discuss, in general terms, how integrating out heavy moduli which break supersymmetry can have strong effects on the effective inflaton potential. We derive explicit formulae for the latter, assuming that the inflaton and moduli sectors interact only gravitationally. For the sake of completeness, this is done in the cases with and without the stabilizer field. We illustrate these general results with three examples -- KKLT moduli stabilization, K\"ahler Uplifting, and the Large Volume Scenario -- in Secs.~3, 4, and 5, respectively. In each model we briefly review the stabilization mechanism itself and provide different techniques which can be used to integrate out the moduli. We derive bounds on the gravitino mass and the field value of the inflaton arising from stability of the moduli. Furthermore, each one of the examples is illustrated by means of a numerical example. In Sec.~6 we discuss the universality of the leading-order effective inflaton potential arising in all our examples and the shared universal CMB observables this predicts. Finally, our results are discussed in Sec.~7, and technical details are summarized in the Appendices A and B.

%
%

\section{Integrating out supersymmetry-breaking moduli}
\label{sec:General}

%

\subsection{Effects of supersymmetry breaking}
\label{sec:General1}

We are interested in supergravity models in which the inflaton field $\varphi$, which is the imaginary part of a complex scalar field $\phi = \frac{1}{\sqrt{2}} (\chi + i \varphi)$, interacts with heavy moduli and supersymmetry breaking fields, collectively denoted by $T_{\alpha}$. The effective action is defined by
\begin{align}\begin{split}
 K &= K_0 (T_{\alpha}, {\overline T}_{\bar \alpha} ) + \frac{1}{2} K_1 (T_{\alpha}, {\overline T}_{ \bar \alpha}) (\phi + {\bar \phi})^2 \,, \\
 W &=  W_\text{mod}  (T_{\alpha}) + \frac12 m \phi^2 \,. \label{g1}
\end{split}\end{align}
It can potentially reconcile chaotic inflation, moduli stabilization, and supersymmetry breaking. We are interested in the regime where the moduli and the supersymmetry breaking fields $T_{\alpha}$ are much heavier than the inflaton. Such heavy fields usually decouple from low-energy dynamics once they settle into their minima, denoted by $T_{\alpha,0}$. The case without supersymmetry breaking was studied in\cite{Buchmuller:2014vda}. It was shown that for a single heavy modulus $T$ with $K_0 (T, \overline T) = - 3 \ln{\left(T + \overline T \right)}$ and $K_1(T, \overline T) = 1$ the effects on the dynamics of inflation can be expressed as 
\begin{align}\begin{split}\label{eq:KLEffPot}
V &\approx \frac{V_\text{inf}(\phi_\alpha)}{(2 T_0)^3} - \frac{3}{2 (2 T_0)^{9/2} \,m_T} \Big\{ W_\text{inf} \Big[V_\text{inf}(\phi_\alpha) + e^K K^{\alpha \bar \alpha} \partial_\alpha W_\text{inf} D_{\bar \alpha} \overline W_\text{inf} \Big] + \text{ c.c.} \Big\} \\ &- \frac{3 e^K}{(2 T_0)^{6}\, m_T^2} \Big|K^{\alpha \bar \alpha} D_\alpha W_\text{inf} \partial_{\bar \alpha} \overline W_\text{inf} \Big|^2\,,
\end{split}\end{align}
up to terms suppressed by higher powers of the modulus mass $m_T$. Here, $W_\text{inf}$ denotes the superpotential of the inflaton sector, comprised of scalar fields $\phi_\alpha$. $V_\text{inf}(\phi_\alpha)$ denotes the inflaton scalar potential in the absence of a modulus sector. Evidently, all corrections stemming from integrating out the heavy modulus disappear in the limit $m_T \to \infty$. 

However, if any of the fields $T_\alpha$ break supersymmetry the picture changes. In this case, there are well-known effects that do not decouple from inflation. In the context of low-energy supersymmetric models these lead to soft-breaking terms whose size is controlled by the gravitino mass. In particular, considering spontaneous supersymmetry breaking we expect the effective inflaton potential to be of the form
\begin{align} 
V = V_\text{SUGRA} + \frac{c}{2}  {\tilde m} m_{3/2} \varphi^2 + \dots \,, \label{g2}
\end{align}
where $c$ is a model-dependent real constant and $V_\text{SUGRA}$ is to be computed using
\begin{align}
K = \frac{1}{2} \left(\phi + {\bar \phi}\right)^2\,, \qquad W =  \frac12 \tilde m \phi^2\,,  \label{g3}
\end{align}
with ${\tilde m} = K_1^{-1}e^{\frac12 K_0 (T_0, {\overline T}_0 )} m$ and the wave-function normalization $\phi \to K_1^{-1/2} \phi$ to match the notation of Eq.~\eqref{g1}. Notice that in Eq.~\eqref{g2} a term proportional to $m_{3/2}^2 \varphi^2$ is absent due to the shift symmetry $\phi \to \phi + i \alpha$, which is broken softly by the mass term in the superpotential. Computing $V_\text{SUGRA}$ from Eqs.~\eqref{g3} while imposing cancellation of the cosmological constant at the end of inflation, $\varphi = 0$, and setting the heavy real scalar $\chi$ to its minimum at $\langle \chi \rangle =0$, we find
\begin{align}
V = \frac{1}{2} {\tilde m}^2 \varphi^2 + \frac{c}{2} {\tilde m} m_{3/2} \varphi^2 - \frac{3}{16} {\tilde m}^2
 \varphi^4 + \dots \,.  \label{g5}
\end{align} 
Apparently, the second term only decouples from inflation if $m_{3/2} \ll \tilde m$. The dots in Eqs.~\eqref{g2} and \eqref{g5} denote sub-leading terms and higher powers in $\varphi$, for example terms of order $\mathcal O( {\tilde m} m_{3/2} \varphi^4)$. Usually, such terms can be discarded easily. In large-field inflation, however, super-Planckian excursions of $\varphi$ can make corrections relevant. Therefore, in the following we systematically calculate corrections to the leading-order potential in Eq.~\eqref{g5}. We are curious to find out if corrections from the modulus sector can cancel the third term in the effective potential, which makes $V$ unbounded from below. Furthermore, if the modulus sector has an approximate no-scale symmetry we expect a cancellation of the bilinear soft mass term, i.e., $c\ll 1$. We wish to discuss if, in this situation, chaotic inflation can proceed via the supersymmetric mass term of $\varphi$ without spoiling the stabilization of moduli.

%

\subsection{Integrating out heavy moduli }
\label{sec:General2}

In the following, we would like to generalize the results of \cite{Buchmuller:2014vda}, in particular Eq.~\eqref{eq:KLEffPot}, to more general supergravity Lagrangians. Starting from Eqs.~\eqref{g1} we find for the K\"ahler metric and its inverse
\begin{align}
K_{I {\bar J}} = \begin{pmatrix} K_{0,\alpha {\bar \beta}} & 0 \\ 0 & K_1 \end{pmatrix} \,, \qquad 
K^{{ I} \bar J} = \begin{pmatrix} K_0^{{ \alpha} \bar \beta} & 0 \\ 0 & K_1^{-1} \end{pmatrix} \,. \label{sp2}
\end{align}
The indices $I$ and $J$ run over the $T_\alpha$ and $\phi$. Accordingly, the scalar potential is given by
\begin{align}\begin{split}
V = e^{K_0}  &\Bigg\{  K_0^{{\alpha} \bar \beta} \left[  
W_{\alpha} +K_{0, \alpha }  \left(W_\text{mod} (T_{\alpha}) + \frac12 m \phi^2\right) \right]
\left[ \overline{{W}}_{\bar \beta} + K_{0, {\bar \beta} } \left( \overline W_\text{mod} (\overline T_{\bar \alpha}) + \frac12 m \bar \phi^2 \right) \right] \\
 &+ K_1^{-1} m^2 |\phi|^2 - 3 \left|W_\text{mod}  (T_{\alpha}) + \frac12 m \phi^2\right|^2 \Bigg\} \,. \label{sp3}
\end{split}\end{align}
Assuming the cosmological constant to be canceled at $\phi = 0$, i.e., after inflation has ended, means 
\begin{align}
K_0^{{ \alpha} \bar \beta} \left[ W_{\alpha} +K_{0, \alpha }  W_\text{mod}  \right]
\left[ \overline{{ W}}_{\bar \beta} +K_{0, {\bar \beta} }  \overline{W}_\text{mod}  \right] = 3 \left|W_\text{mod}  (T_{\alpha,0})\right|^2 \,. \label{sp4}
\end{align}
Furthermore, we assume that the moduli fields adiabatically trace the minimum of their potential during inflation. This is justified as long as their masses are larger than the Hubble scale. Specifically,
\begin{align}
\nabla_{\alpha} V = 0 \qquad \Rightarrow \qquad G^I \nabla_{\alpha} G_I + G_{\alpha} = 0 \,. \label{sp5}
\end{align} 
Here $\nabla_{\alpha}$ denotes the covariant derivative on field space, i.e., $\nabla_{\alpha} G_I = G_{\alpha I} - \Gamma_{\alpha I}^J G_J$ in terms of the K\"ahler function $G = K+ \ln |W|^2$, where $\Gamma$ is defined in Appendix~A.1. 

We can now integrate out the heavy fields $T_\alpha$ to obtain an effective scalar potential for the inflaton $\varphi$. Using Eq.~\eqref{sp4} and using that $\chi$ is heavy due to its soft mass and stabilized at the origin we can expand $V$ in powers of the inflaton field,
\begin{align}
V &= e^{K_0} \Bigg\{ K_0^{{\alpha} \bar \beta} \left[ W_{\alpha} +K_{0, \alpha }  \left(W_\text{mod} (T_{\alpha}) - \frac14 m \varphi^2\right) \right] \left[ \overline{W}_{\bar \beta} + K_{0, {\bar \beta} } \left( \overline W_\text{mod} (\overline T_{\bar \alpha}) - \frac14 m \varphi^2 \right) \right]  \nonumber \\
& \ \ \ \ \ \ \ \ + \frac12 K_1^{-1} m^2 \varphi^2 - 3 \left|W_\text{mod}  (T_{\alpha}) - \frac14 m \varphi^2\right|^2 \Bigg\} \nonumber\\
& = V_0  (T_{\alpha}, {\overline T}_{\bar \alpha} ) + \frac{1}{2} V_1 (T_{\alpha}, {\overline T}_{\bar \alpha})  m \varphi^2 +  \frac{1}{4} V_2 (T_{\alpha}, {\overline T}_{\bar \alpha})  m^2 \varphi^4 \,. \label{sp6}
\end{align}
The explicit coefficients $V_0$, $V_1$, and $V_2$ and other details of the computation are given in Appendix~\ref{app:GeneralExpansion}. During inflation the fields $T_\alpha$ are displaced from their minima, 
\begin{align}
T_{\alpha} \ = \ T_{\alpha,0} + \delta T_{\alpha}\,. \label{sp8}
\end{align}
We can expand the coefficients $V_i$ in Eq.~\eqref{sp6} at leading order in $\delta T_\alpha$ as long as $|\delta T_{\alpha}| \ll  |T_{\alpha,0}|$. Introducing $\rho_{\alpha} = (T_{\alpha},{\overline T}_{\bar \alpha})$ this can be written as 
\begin{align}
V = \frac{1}{2} \delta \rho_{\alpha} M^2_{\alpha \beta}  \delta \rho_{\beta} + \frac{1}{2} \left( V_1 +  \frac{\partial V_1}{\partial \rho_{\alpha}} \delta \rho_{\alpha}\right) m \varphi^2 + \frac14 V_2 m^2 \varphi^4 + \dots \,, \label{sp10}
\end{align}
where $M^2_{\alpha \beta}$ denotes the un-normalized mass matrix of the $\rho_\alpha$. Again, details can be found in Appendix~\ref{app:GeneralExpansion}. Minimizing this expression with respect to $\delta \rho_\alpha$ we find for the displacement of the moduli at leading order,
\begin{align}
\delta \rho_{\alpha} = - \frac{1}{2} (M^{-2})^{\alpha \beta} \frac{\partial V_1}{\partial \rho_{\beta}} m \varphi^2 \,. \label{sp11} 
\end{align}
Plugging this back into Eq.~\eqref{sp10} we obtain the effective potential in its most general form, 
\begin{align}\begin{split}\label{eq:GeneralEffPot}
V &= \frac12 V_1 \left(T_{\alpha,0}, {\overline T}_{\bar \alpha,0}\right) m \varphi^2 
+ \frac14 V_2 \left(T_{\alpha,0}, {\overline T}_{\bar \alpha,0} \right) m^2 \varphi^4
\\ &- \frac12 
\begin{pmatrix} 
\frac{\partial V_1}{\partial T_{\alpha}} & \frac{\partial V_1}{\partial {\overline T}_{\bar \alpha}} 
\end{pmatrix}  
\begin{pmatrix} 
(m^{-2})^{\alpha {\bar \beta}} & (m^{-2})^{\alpha {\beta}}  \\ (m^{-2})^{\bar \alpha {\bar \beta}} & (m^{-2})^{\bar \alpha {\beta}}  
\end{pmatrix} 
\begin{pmatrix} 
\frac{\partial V_1}{\partial {\overline T}_{\bar \beta}} \\ \frac{\partial V_1}{\partial T_{\beta}}  
\end{pmatrix} 
m^2 \varphi^4 + \dots\,.
\end{split}\end{align}

To simplify this expression it is useful to consider a limit in which supersymmetry is weakly broken, cf.~the more detailed discussion in Appendix~A.2. This is the case when the supersymmetric mass, i.e., the mass of the fermions associated with the scalars $T_\alpha$, is much larger than the gravitino mass.\footnote{With the exception of the goldstino, of course.} Specifically, when 
\begin{align}
\text{Eigenvalues}\left[({m_F})_{\alpha \beta}\right] = \text{Eigenvalues}\left[e^{G/2} \left(\nabla_{\alpha} G_{\beta} + \frac{1}{3} G_{\alpha} G_{\beta}\right)\right] \gg m_{3/2}\,.
\end{align}
Alternatively, one may consider the case where the supersymmetry breaking scale is large but the supersymmetry breaking sector decouples from moduli stabilization. An example for this is supersymmetry breaking in the O'Raifeartaigh model with a very heavy Polonyi field. For both of these possibilities the effective inflaton potential becomes
\begin{align}\begin{split}
V &\approx \frac{m \varphi^2}{2}e^{K_0} \left\{ - \frac{1}{2} K_0^{\alpha \bar \beta} \left( K_{0,{\bar \beta}} D_{\alpha} W_\text{mod} + K_{0,\alpha } \overline{D}_{\bar \beta} \overline{W}_\text{mod} \right) + m K_1^{-1} +  \frac{3}{2} (W_\text{mod} + \overline{W}_\text{mod}) \right\} \\
&+ \frac{m^2 \varphi^4}{16} e^{K_0} \bigg\{- 3 + e^{K_0/2} \Big[ K_{\delta} \left(m_F^{-1}\right)^{\beta \delta} \Big[- K_0^{\epsilon \bar \epsilon} (K_{\beta \epsilon} + K_{\beta} K_{\epsilon} - \Gamma_{\beta \epsilon}^{\gamma} K_{\gamma}) \overline{D}_{\bar \epsilon} \overline{W}_\text{mod} \\
&\hspace{2.1cm}+ 2 D_{\beta} W_\text{mod} + 3 K_{\beta} {\overline W}_\text{mod} +2mK_1^{-2}\left(K_{0,\beta} K_1 - K_{1,\beta}  \right)\Big] + \text{h.c.}\Big] \bigg\} \,, \label{sp14} 
\end{split}\end{align}
which is the desired generalization of Eq.~\eqref{eq:KLEffPot}. Notice, however, that the quadratic term is independent of the small-supersymmetry breaking approximation. It is simply the total mass -- supersymmetric and soft mass -- of the inflaton in the true vacuum, computed from the effective action defined by (\ref{g1}). Indeed, using the definition of the inflaton ${\rm Im} \, \phi = \varphi/\sqrt{2}$ and the supergravity scalar masses in Eqs.~(\ref{m1}), we find that the inflaton mass is
\begin{equation}
m_{\varphi}^2 = m_{\phi {\bar \phi}}^2 - \frac{1}{2} \left( m_{\phi \phi}^2 + m_{{\bar \phi} 
{\bar \phi}}^2 \right)  
\ . \label{inflatonmass}
\end{equation}
It is a straight-forward, though non-trivial exercise to prove that Eq.~(\ref{inflatonmass}) equals the mass term in the first line of Eq.~(\ref{sp14}).

Using this result we can, in principle, calculate the effective potential with corrections for any model of moduli stabilization described by the ansatz Eqs.~\eqref{g1}. In practice, however, the approximation outlined above to obtain Eq.~\eqref{sp14} -- more precisely, the quartic term, as explained above -- is not always applicable. In that case, either a more general expression for the effective potential can be used, given by Eq.~\eqref{eq:GeneralEffPot}, or the calculation can be significantly simplified by expanding in small parameters while performing the above analysis. Before demonstrating this in three popular examples of moduli stabilization with spontaneously broken supersymmetry, we give a short remark on chaotic inflation with a stabilizer field.

%

\subsection{Chaotic inflation with a stabilizer field}
\label{sec:General3}

Although the main focus of this paper is the simple chaotic inflation model with a quadratic superpotential, we consider it worthwhile to make a couple of remarks about the scenario with a stabilizer field. This model has been intensively studied in the literature, and its interplay with supersymmetric moduli stabilization has been treated in \cite{Davis:2008fv,Kallosh:2011qk,Buchmuller:2014vda}. A generalization of the results in \cite{Buchmuller:2014vda} can be found analogously to the above analysis.

As a starting point we consider
\begin{align}\begin{split}
K &= K_0 (T_{\alpha}, {\overline T}_{\bar \alpha} ) + \frac{1}{2} K_1 (T_{\alpha}, {\overline T }_{\bar \alpha} ) (\phi + {\bar \phi})^2 + K_{S \bar S} |S|^2 +  \frac{1}{4} K_{S \bar S S \bar S } |S|^4  + \dots\,, \\
W &= W_\text{mod}  (T_{\alpha}) + m S \phi \,. \label{s1}
\end{split}\end{align}
As before, for simplicity we assume the superpotentials of the moduli sector and inflation sector to be decoupled. The canonically normalized inflaton is $ \varphi = \sqrt{2} \text{Im}\, \phi $, and $ \psi = \sqrt{2} \text{Im}\, S$. The real parts of $\phi$ and $S$ are assumed to be stabilized at the origin.
The scalar potential is given by
\begin{align}\begin{split}
V &= e^{K}  \bigg\{  K_0^{{\alpha} \bar \beta} \left[  W_{\alpha} +K_{0, \alpha }  (W_\text{mod} + m S \phi) \right] \left[ \overline{{W}}_{\bar \beta} + K_{0, {\bar \beta} }  (\overline{W}_\text{mod} + \overline{m S \phi}) \right] + \  K_1^{-1} m^2 |S|^2 \\
& \hspace{0.5cm}+ \frac{1}{K_{S \bar S} + K_{S \bar S S \bar S } |S|^2} \left|\left(1 + K_{S \bar S} |S|^2\right) m \phi + K_{S \bar S} {\bar S} W_\text{mod}\right|^2 - 3 \left|W_\text{mod}+ m S \phi\right|^2 \bigg \} \, . \label{s2}
\end{split}\end{align}
Imposing cancellation of the cosmological constant, Eq.~\eqref{sp4}, and stabilization of all $T_\alpha$ and $S$ during inflation,
\begin{align}\begin{split}
\nabla_{\alpha} V &= 0 \qquad \Rightarrow \qquad G^I \nabla_{\alpha} G_I + G_{\alpha} = 0 \,,  \\
\nabla_{S} V &= 0 \qquad \Rightarrow \qquad G^I \nabla_{S} G_I + G_{S} = 0  \, ,  \label{s3}
\end{split}\end{align}  
we can again integrate out the heavy $T_\alpha$. Details of this computation are given in Appendix~\ref{app:Stabilizer}. Expanding in powers of the inflaton we find 
\begin{align}
 V =  V_0  (T_{\alpha}, S, {\overline T }_{\alpha}, \bar S ) + V_1 (T_{\alpha}, S, {\overline T }_{\alpha}, \bar S ) m \psi \varphi +  \frac{1}{2} V_2 (T_{\alpha}, S, {\overline T }_{\alpha}, \bar S) m^2 \varphi^2 \,.\label{s4}
\end{align}
Expanding this in the moduli displacements and in $\delta \psi$ during inflation, with $\langle S \rangle =0$, yields
\begin{align}\begin{split}
V &= \frac{1}{2} \delta \rho_{\alpha} M^2_{\alpha \beta}  \delta \rho_{\beta}  + \frac12 \delta \psi^2 \left[ m_S^2 + \frac12 m^2 \varphi^2 e^{K_0} \left( K_0^{\alpha \bar \beta} K_{0, \alpha }  K_{0, {\bar \beta} } - \frac{K_{S \bar S S \bar S} }{K_{S \bar S }^2 } \right) \right] \\
&\hspace{0.3cm} + m \varphi \delta \psi \left( V_1 +  \frac{\partial V_1}{\partial \rho_{\alpha}} \delta \rho_{\alpha} \right) + \frac12 m^2 \varphi^2 \frac{e^{K_0}}{ K_{S \bar S}}  + \dots \,, \label{s7}
\end{split}\end{align}
with $\rho_\alpha = (T_\alpha,{\overline T }_{\bar \alpha})$. Consequently, 
\begin{align}
\delta \rho_{\alpha} = - m \varphi (M^{-2})^{\alpha \beta} \frac{\partial V_1}{\partial \rho_{\beta}} \delta \psi\,, \label{s8}
\end{align}
with
\begin{align}
\delta \psi = - \frac{m \varphi V_1} { m_S^2 + \frac12 m^2 \varphi^2 e^{K_0} \left( K_0^{\alpha \bar \beta} K_{0 \alpha } K_{0 {\bar \beta} } - \frac{K_{S \bar S S \bar S} }{K_{S \bar S}^2 } \right) } \,. \label{ss8}
\end{align}

In the near-supersymmetric limit outlined in Appendix~\ref{app:Massapprox} we find for the effective inflaton potential
\begin{align}\label{s11}
V &\approx \frac12 m^2 \varphi^2 \left( K_{S \bar S}^{-1} e^{K_0} - \frac{V_1^2}{m_S^2} \right)  - \frac{ V_1^2 e^{K_0}}{4 m_S^4} m^4  \varphi^4 \Bigg\{ \frac{K_{S \bar S S \bar S}}{ K_{S \bar S}^{2}} 
+ e^{K_0/2} \bigg[   K_{\delta} (m_F^{-1})^{\beta \delta} \\ &\hspace{0.6cm}\times \Big[K_0^{\epsilon \bar \epsilon} (K_{\beta \epsilon} + K_{\beta} K_{\epsilon} - \Gamma_{\beta \epsilon}^{\gamma} K_{\gamma}) \overline{D}_{\bar \epsilon} \overline{W}_\text{mod} - D_{\beta} W_\text{mod} - \frac{1}{2} K_{\beta} {\overline W_\text{mod}} \Big] + \text{h.c.}  \bigg] \Bigg\}\,, \nonumber
\end{align}
where 
\begin{align}
V_1=V_1\Big|_{S=0} = - \frac{1}{2} e^{K}  \bigg\{  K_0^{\alpha \bar \beta} \left( K_{0, {\bar \beta}} D_{\alpha} W_\text{mod} + K_{0, \alpha } \overline{D}_{\bar \beta} \overline {W}_\text{mod} \right)  - 2 (W_\text{mod} + \overline{W}_\text{mod}) \bigg\} \,.
\end{align}
Analogous to the case without stabilizer the quadratic term in $\varphi$ is independent of the small-supersymmetry breaking approximation.

Let us compare this result to the the case without stabilizer, Eq.~\eqref{sp14}. Since $V_1 \sim m_{3/2}$ and $m_S^2 \sim \tilde m^2 + m_{3/2}^2$, the corrections to the chaotic scalar potential $\frac12 {\tilde m}^2 \varphi^2$, with $\tilde m = m e^{K_0/2}$, are negligible for $m_{3/2} \ll {\tilde m}$. 
For large gravitino masses $m_{3/2} \gtrsim {\tilde m}$, on the other hand, the quadratic inflaton term in Eq.~\eqref{s11} becomes negative and stops inflation. Simultaneously, the quartic term becomes sizeable. 
Thus, these generic results fit nicely with the explicit analysis performed in \cite{Buchmuller:2014pla}. However, remember that Eq.~\eqref{s11} is only valid in the near-supersymmetric limit. If the supersymmetry-breaking $T_\alpha$ can not be completely decoupled in the fermion mass matrix, the appropriate quartic term in the scalar potential is given by the more general result Eq.~\eqref{sss8}. Since all moduli stabilization schemes with supersymmetry breaking that we consider require a large gravitino mass, it is difficult to reconcile these schemes with chaotic inflation with a stabilizer. Therefore, in the examples treated in the following sections we restrict ourselves to the more interesting models with no stabilizer field.

%

\section{Chaotic inflation with KKLT moduli stabilization}
\label{sec:KKLT}

As a first example we discuss stabilization of a single K\"ahler modulus $T$ by the mechanism of KKLT\cite{Kachru:2003aw} and its interaction with chaotic inflation. Before treating the coupled Lagrangian we discuss important properties of the original KKLT vacuum and its uplift. Many of these are well-known facts, nonetheless it is instructive to review them before discussing the interaction with inflation.

%

\subsection{KKLT moduli stabilization and uplift}
\label{sec:KKLT1}

The possibly simplest setup to stabilize K\"ahler moduli via non-perturbative effects was proposed in \cite{Kachru:2003aw}. The original model assumes all complex structure moduli of a compact CY manifold and the dilaton to be stabilized by fluxes, as first developed in \cite{Giddings:2001yu}. The remaining effective theory contains a single lightest K\"ahler modulus, in the following denoted by $T$, which parameterizes the volume of the compact manifold. $T$ then has the following tree-level K\"ahler potential,
\begin{align}
K = - 3 \ln{\left(T + \overline T \right)}\,,
\end{align}
and does not appear in the flux superpotential, $W_0$, responsible for stabilizing the complex structure and the dilaton. Therefore, $T$ is massless at perturbative tree-level and must be stabilized to avoid a series of well-known problems. This is achieved by employing non-perturbative corrections to the superpotential, so that $W$ takes the form
\begin{align}
W = W_0 + A e^{- a T}\,.
\end{align}
We treat $W_0$ and $A$ as constants determined by fluxes and vacuum expectation values of complex structure moduli. They are assumed to be real in what follows. A relative phase between $A$ and $W_0$ can always be compensated by a field redefinition. Depending on whether the non-perturbative term stems from a Euclidean D3 instanton or from a gaugino condensate on a stack of D7 branes, $a$ can be $2 \pi$ or $\frac{2 \pi}{N}$, where $N$ is the rank of the condensing gauge group. The scalar potential 
\begin{align}\label{V}
V = e^K \left( K^{T \overline T}D_T W \overline{D_T W} - 3 |W|^2 \right)\,,
\end{align}
has two extrema, $\partial_T V = 0$, corresponding to
\begin{align}\label{DTAdS}
D_T W = 0\,.
\end{align}
One extremum lies at $T=\infty$, where the potential vanishes. In addition there is
a supersymmetric AdS vacuum at $\tilde T_0$  which is determined by
\begin{align}\label{eq:KKLTvac}
W_0 = -A e^{-a \tilde T_0} \left(1+\frac23 a \tilde T_0 \right)\,.
\end{align}
For real parameters of the superpotential $\tilde T_0 $ is real. $\text{Im}\, T$ is stabilized at the origin at the same mass scale as $\text{Re}\,T$. 

To uplift the AdS vacuum to a Minkowski vacuum the authors of \cite{Kachru:2003aw} introduced an anti-D3 brane. To avoid explicit supersymmetry breaking\footnote{See, however, \cite{Kallosh:2014wsa} for a very recent treatment of this issue.} we resort to uplifting via the F-term of a Polonyi field $X$, with 
\begin{align}
K_\text{up} = k\!\left(|X|^2\right)\,, \qquad W_\text{up} = f X\,.
\end{align}
Uplifting of AdS vacua via F-terms of matter fields was first discussed in \cite{Lebedev:2006qq}. We assume that the function $k$ contains a quartic term so that $X$ is stabilized close to the origin at a high scale, and thus the field completely decouples from the dynamics of moduli stabilization and inflation. Such a quartic term may effectively arise from couplings to heavy fields, cf.~\cite{O'Raifeartaigh:1975pr}.\footnote{For a more thorough treatment of the dynamics linking supersymmetry breaking and chaotic inflation, cf.~\cite{Kallosh:2011qk,Buchmuller:2014pla}.} The only contribution of the Polonyi field to $V$ is then its F-term,
\begin{align}\label{Vup}
V_\text{up} = e^K f^2\,,
\end{align}
which can be used to cancel the cosmological constant in the true vacuum defined by Eq.~\eqref{eq:KKLTvac}. 

In addition to the extremum at $T = \infty$ corresponding to $D_T W = 0$, the uplifted scalar potential has two further extrema which are determined by
\begin{align}\label{DTupKKLT}
D_T W = - \frac{3 W}{4 T} \left( 1 \pm \sqrt{1- \frac{2f^2}{ (a T +2) W^2}}\right)\,.
\end{align}
The negative sign yields the uplifted AdS minimum,
\begin{align}\label{DTAdSup}
D_T W = -\frac{3f^2}{4 a T_0^2 \left. W\right|_{T_0} } + \mathcal{O}(T_0^{-3})\,,
\end{align}
where the value of the modulus $T$ is shifted to $T_0 = \tilde T_0 + \delta T_\text{up}$. The shift in $T$ is easily obtained by expanding $D_T W$ in $\delta T_\text{up}$,
\begin{align}\label{method}
\left. D_T W\right|_{\tilde T_0} &\approx \left. D_T W\right|_{T_0} - \delta T_\text{up} \left.\partial_T D_T W\right|_{T_0} 
\nonumber\\
&\approx \left. D_T W\right|_{T_0} - \delta T_\text{up}\left( \left(-a + K_T\right) \left. D_T W\right|_{T_0}
+ \left( \left( a - K_T\right)K_T  + \partial_T K_T\right)\left. W\right|_{T_0} \right) \,.
\end{align}
Using Eqs.~\eqref{DTAdS} and \eqref{DTAdSup} we find
\begin{align}
\frac{\delta T_\text{up}}{T_0} \approx \frac{f^2}{2 a^2 T_0 W_0^2} + \mathcal O ( T_0^{-2} ) \,,
\end{align}
where we have used $\left. W\right|_{T_0} \approx W_0$. Using Eqs.~\eqref{V}, \eqref{Vup}, and \eqref{DTAdSup} one finds that the cosmological constant of the AdS vacuum is canceled
by tuning $f$ to
\begin{align}
f = \sqrt 3 W_0 \left(1 - \frac{3}{2 a T_0} + \mathcal O (T_0^{-2})\right)\,.
\end{align}
Note that there is a sub-leading contribution of the modulus to supersymmetry breaking,
\begin{align}
\langle F_T \rangle = e^{K/2} \sqrt{K^{T \overline T}} D_T W\Big|_{T_0} \approx - \frac{3 \sqrt3 W_0}{a (2T_0)^{5/2}} \approx 
- \frac{3\langle F_X \rangle}{4a T_0}\,.
\end{align}
Since $a T_0 \gg 1$ for consistency of the single-instanton approximation, the dominant contribution to supersymmetry breaking stems from the Polonyi field. The gravitino mass in the Minkowski vacuum is given by
\begin{align}\label{eq:KKLTm32}
m_{3/2} = e^{K/2}W = \frac{W_0}{(2 T_0)^{3/2}} \left(1 - \frac{3}{2 a T_0} + \mathcal O(a T_0)^{-2} \right) \approx \frac{W_0}{(2 T_0)^{3/2}} \,.
\end{align}
It is closely related to the mass of the canonically normalized modulus,
\begin{align}\label{eq:KKLTMass}
m_T \approx 2 a T_0 m_{3/2}\,.
\end{align}

The uplifted Minkowski vacuum is protected by a barrier from the run-away vacuum at $T = \infty$. The height of the barrier can be found by choosing the positive sign in the expression \eqref{DTupKKLT} for the covariant derivative, corresponding to the local maximum in the scalar potential. For the field value of the modulus at the position of the barrier, $T_\text B$, we find
\begin{align}\label{eq:KKLTBarrier}
V_\text{B} = V\Big|_{T_\text B} \approx \frac{f^2}{(2T_\text B)^{3}} \sim 3 m_{3/2}^2\,.
\end{align}

We are now ready to analyze the effect of chaotic inflation on the uplifted KKLT vacuum. Since the F-term of $T$ does not vanish, one may hope that it can cure the problem of unboundedness which plagues the simplest variant of chaotic inflation. To analyze the two-field system defined by the modulus and the inflaton, it is instructive to use both an analytic and a numerical approach.

%

\subsection{KKLT and chaotic inflation: analytic approach}
\label{sec:AnalyticKKLT}

Treating the interaction between the modulus and inflaton sectors in the simplest way, we assume that their superpotentials and K\"ahler potentials completely decouple. Thus, the theory is defined by
\begin{subequations}\label{eq:WKKLT}\begin{align}
W &= W_0 + A e^{-a T} + f X + \frac12 m \phi^2\,, \\
K &= -3 \ln{ \left(T + \overline T \right) + k\!\left(|X|^2\right) + \frac12 \left( \phi + \bar \phi \right)^2}\,.
\end{align}\end{subequations}
In particular, in the notation of Sec.~\ref{sec:General} we choose 
\begin{align}
W_\text{mod}(T_\alpha) &= W_0 + A e^{-a T} + f X \,, \\
K_0(T_\alpha,\overline T_{\bar \alpha}) &= -3 \ln{ \left(T + \overline T \right) + k\!\left(|X|^2\right)}\, \\
K_1(T_\alpha,\overline T_{\bar \alpha}) &= 1\,.
\end{align}
Note that the relative phase between $W_0$ and $m$ is physical. In the following we choose all superpotential parameters to be real, so that only the real part of $T$ is affected by inflation. Therefore, we set $T = \overline T$ in the following discussion. Our results do not change qualitatively if we allow for $m$ and/or $W_0$ to be complex. Moreover, the Polonyi field $X$ is treated in the way discussed in Sec.~\ref{sec:KKLT1}. The canonically normalized inflaton field is $\sqrt 2 \text{Im}\,\phi \equiv \varphi$, which does not appear in the K\"ahler potential. 
On the inflationary trajectory the superpotential reads
\begin{align}\label{superinf}
W = W_0 + A e^{-a T} - \frac{1}{4} m \varphi^2\,.
\end{align}
A natural question to ask is the following: can the effective theory of inflation defined by Eqs.~\eqref{eq:WKKLT} resemble chaotic inflation, after integrating out $T$ at a high scale?

\paragraph{Leading-order effective potential}
$\,$ \\
To answer this question we solve the equation of motion for $T$ during inflation, $\partial_T V = 0$, which yields for the covariant derivative
\begin{align}\label{eq:DTWKKLT}
D_T W = - \frac{3 W}{4 T} \left[ 1 \pm \sqrt{1- \frac{2}{ (a T +2) W^2} \left( f^2 + \frac12 m^2 \varphi^2 \right)} \right]\,,
\end{align}
which implicitly determines $T$ as function of $\varphi$. In addition, there is the extremum at $T = \infty$ with $D_T W = V = 0$. The negative sign in Eq.~\eqref{eq:DTWKKLT} again yields the uplifted AdS minimum, 
\begin{align}\label{DTAdSphi}
D_T W = -\frac{3}{4 a T^2} \frac{f^2 + \frac12 m^2 \varphi^2}{W}+ \mathcal O (T^{-3})\,.
\end{align}
Using Eqs.~\eqref{superinf} and \eqref{DTAdSphi} we obtain for the effective inflaton potential
\begin{align}
V (\varphi) &= \frac{1}{(2 T)^3} \left( f^2 + \frac12 m^2 \varphi^2 - 3W^2 + \mathcal O (T^{-2}) \right) \nonumber \\
 &= \frac12 \tilde m^2 \varphi^2 + \frac32 \tilde m m_{3/2} \varphi^2 - \frac{3}{16} \tilde m^2 \varphi^4 + 
\mathcal O \left(\frac{\delta T}{T_0}\right)\,,\label{eq:EffPotKKLT}
\end{align}
with $\tilde m = \frac{m}{(2 T_0)^{3/2}}$ and $m_{3/2}$ given by Eq.~\eqref{eq:KKLTm32}. The corrections of order $\delta T / T_0$ are due to the $\varphi$-dependent shift of the modulus, $\delta T(\varphi) = T(\varphi) - T_0$. Thus, it seems that after integrating out $T$ the negative definite term proportional to $\tilde m^2 \varphi^4$ still appears in the potential, making it unbounded from below. This is related to the fact that the modulus is only a sub-leading source of supersymmetry breaking. Notice that this way of obtaining the leading-order potential is equivalent to the naive treatment outlined in Sec.~\ref{sec:General1}, which resulted in Eq.~\eqref{g5}.

However, things are not quite as they seem by merely studying the result Eq.~\eqref{eq:EffPotKKLT}. For large values of $\varphi$, i.e., when the quartic term in the effective potential dominates, the modulus can be destabilized by the potential energy of $\varphi$. In this case, the inflationary trajectory becomes tachyonic and the modulus can no longer be integrated out. To see when this point is reached, it suffices to consider the 
structure of Eq.~\eqref{eq:DTWKKLT}. A necessary condition for the existence of real solutions for $D_T W$ is clearly $W^2 \gtrsim 0$. For $W^2 \approx 0$, the uplifted AdS minimum and the maximum merge in a saddle point. Using Eq.~\eqref{superinf} we then obtain an upper bound on allowed values of $\varphi$,
\begin{align}\label{saddlebound}
\tilde m \varphi^2 \lesssim 4m_{3/2}\,.
\end{align} 
This is the well-known bound $H < m_{3/2}$ stressed in \cite{Kallosh:2004yh}, as will become more clear in our numerical example in Sec.~\ref{sec:NumericalKKLT}. There, a more detailed analysis reveals that the modulus is destabilized slightly before the above bound is saturated. In fact, the local maximum of the effective inflaton potential Eq.~\eqref{eq:EffPotKKLT} is never reached while the modulus is stabilized.\footnote{In fact, the full potential defined by Eqs.~\eqref{eq:WKKLT} is bounded from below at all points in field space.}

\paragraph{Corrections to the effective potential}
$\,$ \\
The corrections to the effective potential are determined by the shift of the modulus field $\delta T(\varphi) = T(\varphi) - T_0$.\footnote{Notice that for real superpotential parameters the displacement of $T$ is real as well.} Expanding the covariant derivative in $\delta T$ and
$\varphi^2$, analogous to Eq.~\eqref{method}, we find 
\begin{align}\label{eq:deltaTKKLT}
 \frac{\delta T}{T_0}  = \frac{\tilde m \varphi^2}{ 4 a T_0 m_{3/2}} + \mathcal{O}(T_0^{-2}) \,.
\end{align}
With this, the effective inflaton potential including the leading-order correction becomes, at quartic order in $\varphi$ and leading order in $(a T_0)^{-1}$ and $\tilde m / m_{3/2}$
\begin{align}\label{eq:EffPotKKLT2}
V(\varphi) = \frac12 \tilde m^2 \varphi^2 + \frac32 \tilde m m_{3/2} \varphi^2 - \frac{3}{16} \tilde m^2 \varphi^4 - \frac{3}{4 a T_0} \left(3 \tilde m m_{3/2} \varphi^2  + \frac{3}{4} \tilde m^2 \varphi^4  \right) + \dots \,.
\end{align}
To obtain higher-order corrections to the potential, the potential must be expanded to higher orders in $\delta T$, and $\delta T$ must be computed up to higher powers in $T_0^{-1}$.

So far we have analyzed the deformation of the Minkowski vacuum due to the inflaton field starting from the covariant derivative. Alternatively, on can directly find the shift $\delta T(\varphi)$ by minimizing the scalar potential,
\begin{align}\label{eq:KUTaylor}
V = V|_{T_0} + (\partial_{T} V)|_{T_0} \delta T + \frac12 (\partial_{T}^2 V)|_{T_0} \delta T^2 + \mathcal O(\delta T^3) \,,
\end{align}
along the lines of the general analysis in Sec.~\ref{sec:General2}. One then expects that the shift $\delta T$ is inversely proportional to the modulus mass, cf. Eq.~\eqref{sp11}. Eq.~\eqref{eq:deltaTKKLT} can indeed be rewritten in this form,
\begin{align}\label{KKLTshift2}
\frac{\delta T}{T_0} = \frac{\tilde m \varphi^2}{2 m_T} + \mathcal O(T_0^{-2})\,.
\end{align}

In a manner similar to integrating out $T$, it is possible to verify that the displacement $\delta X$ of the Polonyi field during inflation gives negligible contributions to the inflaton potential. For the particular choice 
\begin{align}
k\left(|X|^2\right) = |X|^2 - \frac{|X|^4}{\Lambda^2}\,,
\end{align}
for example, the displacement of $X$ is at leading order
\begin{align}\label{eq:ShiftX}
\delta X = \Lambda^2 \delta T \,.
\end{align}
Since $\Lambda \ll 1$ to stabilize $X$ at a high scale with a small vacuum expectation value, the contribution of integrating out $X$ at Eq.~\eqref{eq:ShiftX} is clearly negligible.

Among other things, this means that the sector which dominates supersymmetry breaking can be completely decoupled from the dynamics of inflation. In this case, it is possible to obtain the effective potential Eq.~\eqref{eq:EffPotKKLT2} essentially by applying the general expression Eq.~\eqref{sp14}. Details of this computation can be found in Appendix~\ref{app:KKLTExample}.

%

\subsection{A numerical example}
\label{sec:NumericalKKLT}

\begin{figure}[t]
\centering
\includegraphics[width=1\textwidth]{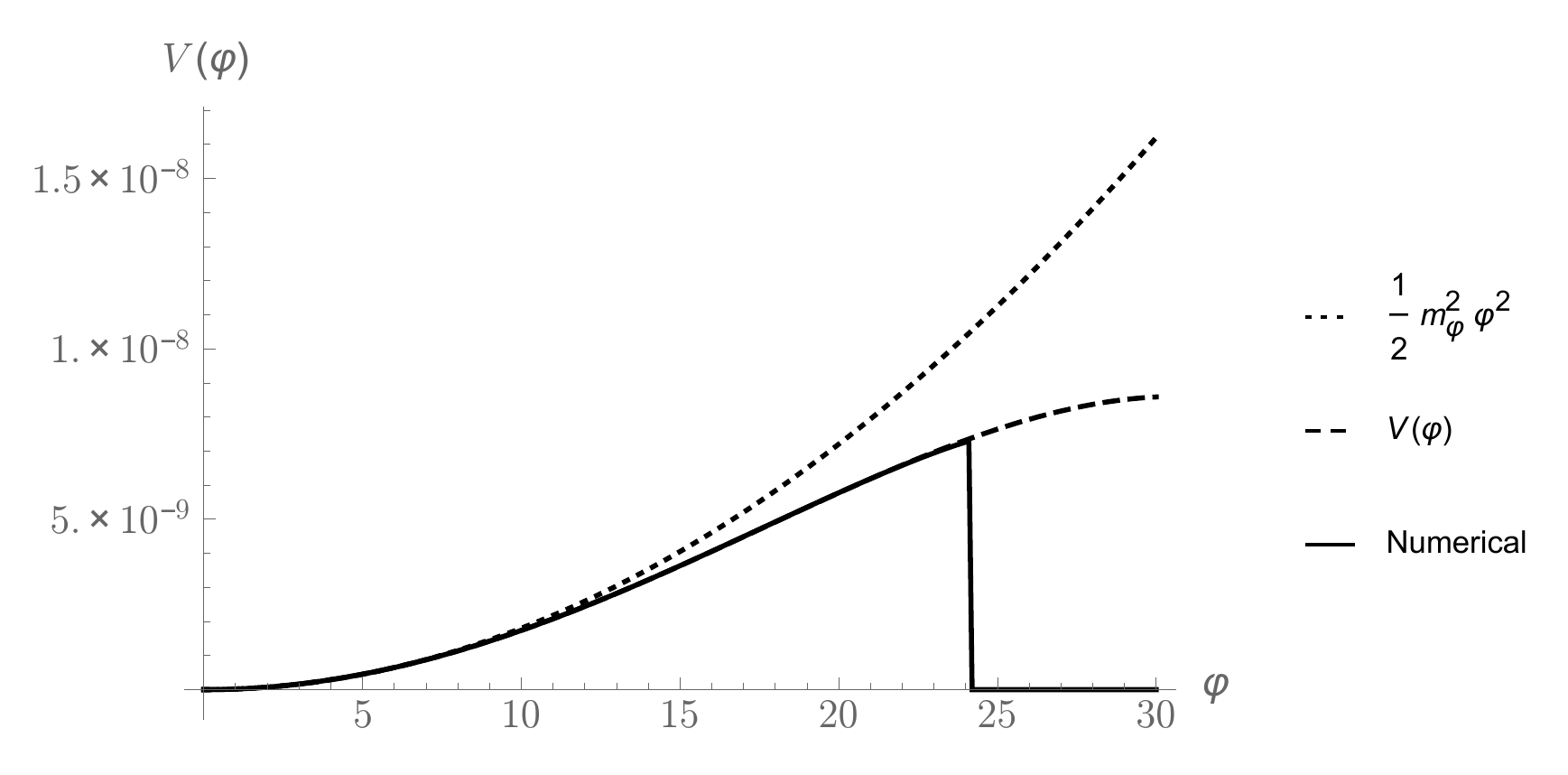} 
\caption{Effective inflaton potential in KKLT for $W_0 = 0.009$, $A=-0.75$, $a = \frac{2 \pi}{10}$, and $m=1.67 \times 10^{-5}$. With these parameters we find $T_0 = 10$ and $m_{3/2} = 10^{-4}$. The dotted line denotes a purely quadratic potential with $m_\varphi= 6 \times 10^{-6}$ imposed by COBE normalization. The dashed line is the effective potential Eq.~\eqref{eq:EffPotKKLT2} evaluated at all orders in $(a T_0)^{-1}$. This potential is valid only as long as the modulus remains stabilized. The solid line is obtained numerically by setting the modulus to its minimum value at each value of $\varphi$. Evidently, above the critical value $\varphi_\text c \approx 24$ the modulus is destabilized towards the run-away minimum at $T=\infty$ and the theory can not be described by Eq.~\eqref{eq:EffPotKKLT2} any longer. \label{fig:PlotKKLT}}
\end{figure}

Let us now study whether 60 e-folds of inflation can be realized with the effective inflaton potential Eq.~\eqref{eq:EffPotKKLT2}, and if the resulting predictions for the CMB observables resemble those of chaotic inflation. It is worth noting that in the parameter regime where $T$ is stabilized, i.e., when $m_{3/2}$ is very large, the bilinear term proportional to $\tilde m m_{3/2}$ actually dominates in $V$ and drives inflation. In this case, the relevant terms in the inflaton potential are 
\begin{align}\label{eq:KKLTpot1}
V(\varphi) \approx \frac32 \tilde m m_{3/2} \varphi^2 \left(1 - \frac{1}{8} \frac{\tilde m}{m_{3/2}} \varphi^2 \right) \,.
\end{align}
Consequently, inflation is only possible if $\tilde m$ and $m_{3/2}$ have the same sign. With Eq.~\eqref{eq:KKLTBarrier} the corrections can be interpreted as a power series in $\frac{H^2}{V_\text{B}}$, the squared Hubble scale divided by the barrier height of the modulus potential. This is a natural expansion parameter because the modulus is destabilized when the vacuum energy of $\varphi$ lifts the modulus over the barrier, cf.~the bound found in Eq.~\eqref{saddlebound}. Neglecting order-one coefficients, COBE normalization imposes ${\sqrt{|\tilde m m_{3/2}|} \sim 3 \times 10^{-6}}$. This puts a lower bound on the gravitino mass, i.e., 
\begin{align}\label{Eq:KKLTGravBound}
m_{3/2} > \sqrt{|\tilde m m_{3/2}|} \, \varphi_\star \sim 5 \times 10^{-5} \sim H\,,
\end{align}
where $\varphi_\star \approx 15$ denotes the inflaton field value at the beginning of the last 60 e-folds of inflation. This means that the gravitino must be very heavy and there is a moderate hierarchy between the  gravitino and inflaton mass for 60 e-folds of chaotic inflation to be possible. This is illustrated in Fig.~\ref{fig:PlotKKLT} for a suitable set of parameters. 

Indeed, 60 e-folds of inflation can take place starting at $\varphi_\star \approx 15$. The CMB observables in our example are found to be 
\begin{align}\begin{split}
n_\text s &= 0.966\,, \\
r &= 0.106\,,
\end{split}\end{align}
which are slightly below the predictions of pure quadratic inflation. This is due to the flattening of the quadratic potential by the negative quartic term. Notice that the modulus is destabilized and the inflaton trajectory becomes tachyonic at the critical value $\varphi_\text c \approx 24$, corresponding to the bound in \eqref{saddlebound}. Therefore, Eq.~\eqref{eq:KKLTpot1} and the dashed line in Fig.~\ref{fig:PlotKKLT} are only meaningful up to this point.

\begin{figure}[t]
\centering
\includegraphics[width=1\textwidth]{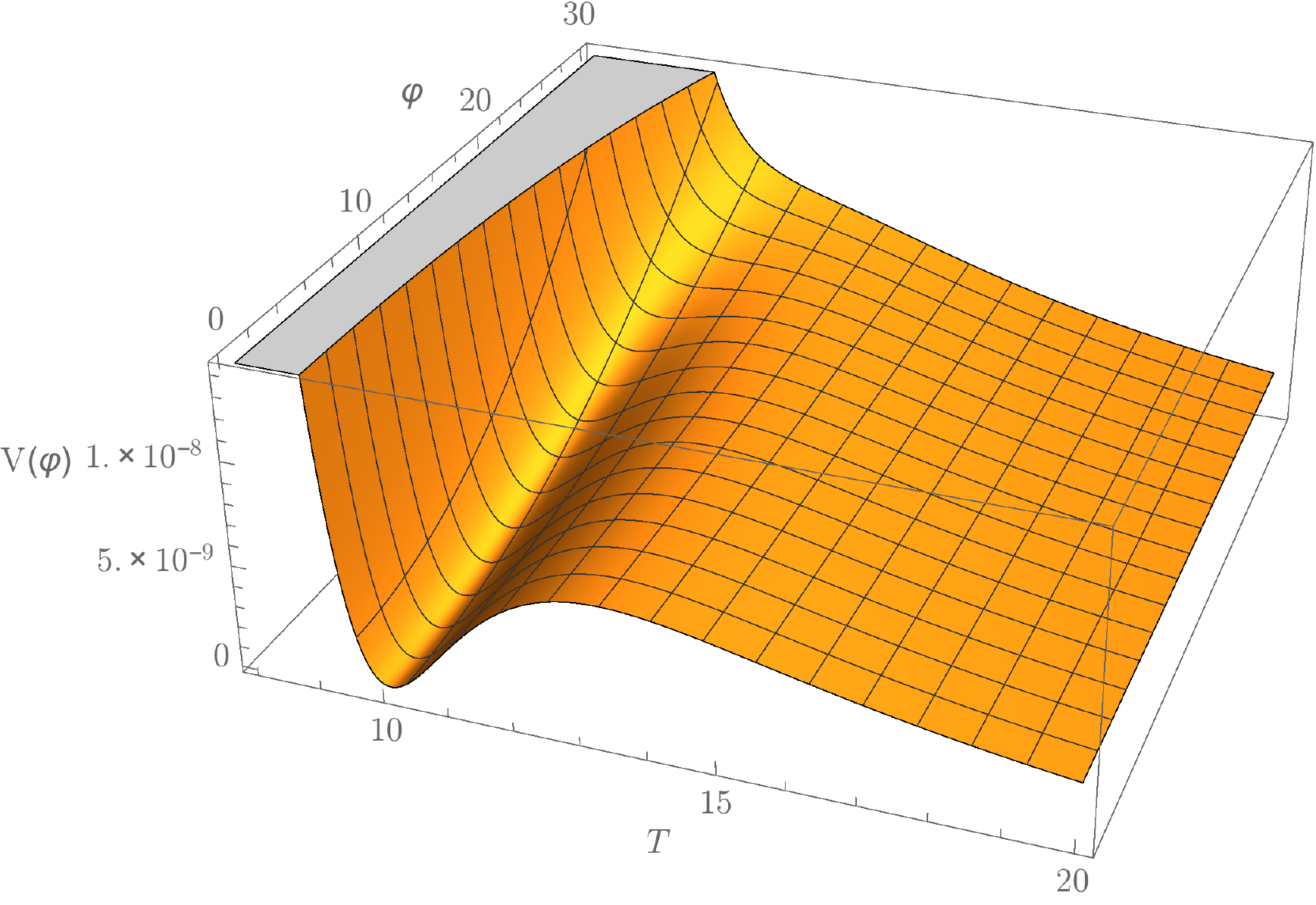} 
\caption{Scalar potential as defined by Eqs.~\eqref{eq:WKKLT} as a function of $T$ and $\varphi$, for the same parameter example as in Fig.~\ref{fig:PlotKKLT}. Apparently, a minimum for the modulus exists for $\varphi \lesssim \varphi_\text c \approx 24$. Beyond this point the modulus runs away towards $T = \infty$ and can no longer be integrated out. For $\varphi < \varphi_\text c$ inflation may take place in the valley of the uplifted modulus minimum. 
\label{fig:Plot3DKKLT}}
\end{figure}

Moreover, the interplay between inflaton and modulus can be illustrated by means of the full scalar potential as a function of $T$ and $\varphi$, depicted in Fig.~\ref{fig:Plot3DKKLT}. The minimum in the modulus direction is uplifted as $\varphi$ increases, until the point where it disappears at $\varphi_\text c \approx 24$.

%

\section{Chaotic inflation with K\"ahler Uplifting}
\label{sec:KU}

%

\subsection{Moduli stabilization by K\"ahler Uplifting}
Another instructive example for moduli stabilization with broken supersymmetry is K\"ahler Uplifting, first proposed in \cite{Balasubramanian:2004uy,Westphal:2006tn}. An appealing feature of this scheme is that K\"ahler moduli can be stabilized in Minkowski or dS vacua without the need of an uplift sector. It is based on the observation that the interplay between a non-perturbative term and a constant term in the superpotential and the leading-order $\alpha'$-correction in the K\"ahler potential can produce local minima in the scalar potential with both negative and positive cosmological constant. In particular, for a careful choice of parameters the Lagrangian defined by 
\begin{align}
W = W_0 + A e^{- a T}\,,
\end{align}
and 
\begin{align}
K = -2 \ln{\left[ \left(T + \overline T \right)^{3/2} + \xi \right]}\,,
\end{align}
can stabilize $T$ in a suitable Minkowski vacuum. Here, $\xi = -\frac{\zeta(3)}{4 (2\pi)^3} \chi \langle \text{Re}\, S\rangle^{3/2}$ where $\chi$ denotes the Euler number of the compactification manifold and $S$ denotes the dilaton. Throughout this work we assume the dilaton to be stabilized supersymmetrically at a high scale so that $\xi$ can be treated as a constant. We remark that this mechanism only works if $\xi$ is positive, hence we only consider negative Euler numbers.

The vacuum structure of this model can again be analyzed by means of the covariant derivative. The extrema of the potential, found by solving $\partial_T V = 0$, correspond to
\begin{align}\label{DTKU}
D_T W = 0 \qquad \text{or} \qquad D_T W = Y W \,,
\end{align}
where the function $Y(T,\overline T)$ is given in Appendix~\ref{app:KU}. The second equation is of particular interest because it allows a vacuum with vanishing cosmological constant, i.e.,
\begin{align}\label{Mink}
D_T W = \pm \sqrt 3 K_{T\overline T}^{1/2} W \,.
\end{align}
Together with Eq.~\eqref{DTKU} this yields
\begin{align}\label{vacuum}
\pm \sqrt 3 K_{T\overline T}^{1/2} = Y \,.
\end{align}
For the negative sign this equation has a solution at large $T_0$ corresponding to $\eta_0 \ll 1$, where we have defined
$\eta = \frac{\xi}{2 (2 T)^{3/2}}$ and $\eta_0 = \eta(T_0)$. Expanding both sides of Eq.~\eqref{vacuum} in powers of $\eta$, cf.~Eqs.~\eqref{KU6} and \eqref{KU7}, we find
\begin{align}\label{eq:KUVac2}
 a T_0 = \frac{5}{2 } - \frac{27 \eta_0}{8} + \mathcal O(\eta_0^{2})\,,
\end{align}
i.e., the vacuum expectation value of the modulus only depends on $a$ and $\xi$.\footnote{Notice that the numerical value $a T_0 \approx 2.5$ is at the border of control over the single-instanton approximation.} A relation between the parameters $W_0$ and $A$ of the superpotential is then obtained from Eq.~\eqref{Mink}, which yields
\begin{align}\label{eq:KUVac1}
 W_0 = -\frac{4}{3 \eta_0} a T_0 A e^{- a T_0} - \frac13 A e^{-a T_0}(3 + 7 a T_0) + \mathcal O(\eta_0)\,.
 \end{align}
Since $\eta_0 \ll 1$ it follows $W_0 \gg A$, contrary to the KKLT case. Therefore, similar to KKLT, the superpotential in the vacuum is dominated by the constant, $W(T) \approx W_0$. Clearly, $T_0$ breaks supersymmetry and the gravitino mass is given by
\begin{align}\label{gravitinoKU}
m_{3/2} = \frac{W_0}{(2 T_0)^{3/2}} \left( 1- \frac{23 \eta_0}{10} + \mathcal O(\eta_0)^2 \right) \approx \frac{W_0}{(2 T_0)^{3/2}}\,.
\end{align}
The extremum with vanishing cosmological constant is a local minimum of the modulus potential. 
The canonically normalized real and imaginary parts of $T$ have the following masses,
\begin{align}\label{eq:KUMass}
m_{\text{Re}\, T}^2 = 5 m_{3/2}^2 \eta_0 + \mathcal O(\eta_0^2)\,, \quad m_{\text{Im}\, T}^2 = \frac{25}{2} m_{3/2}^2 \eta_0 + \mathcal O(\eta_0^2)\,,
\end{align}
respectively. Hence, this particular vacuum disappears if $\xi \to 0$. 

As in KKLT, the potential has an extremum at $T=\infty$ with $D_T W = \partial_T V = V = 0$. Hence, there exists a local maximum at $T_\text B$ with 
\begin{align}\label{eq:KUBarrier}
V_B =e^K(Y^2 - 3)|W|^2\Big|_{T_\text B} \sim \eta_0 m_{3/2}^2\,.
\end{align}
Thus, compared to KKLT, the barrier which separates the Minkowski vacuum from the run-away vacuum is suppressed by a factor $\eta_0$.

Furthermore, the model possesses an AdS minimum at a small value $T_{\rm AdS} \ll \xi^{2/3}$. Although this minimum is not viable from the point of view of supergravity, it is instructive to study its properties in order to understand the differences between KKLT and K\"ahler Uplifting. The supersymmetric AdS minimum corresponds to a solution of $D_T W = 0$. For small $T$ we can perform an expansion in powers of $\xi^{-1}$ which yields, cf.~Eq.~\eqref{KU9},
\begin{align}\label{KUAdS}
\partial_T W\Big|_{T_{\rm AdS}} = -a A e^{-aT_{\rm AdS}} = \left.-K_T W\right|_{T_{\rm AdS}} \approx
\frac{3(2T_{\rm AdS})^{1/2}}{2\xi} W_0 \,.
\end{align}
Since $W_0 \gg A$, this implies $T_{\rm AdS} \ll \xi^2$. The AdS minimum is protected by another barrier, located at $\tilde T_\text{B}$ with $a( \tilde T_\text{B} - T_{\rm AdS}) \approx \ln 2$. The AdS minimum is much deeper than the local Minkowski vacuum, in the sense that its barrier is taller by a factor $\eta_0^{-3}$. Finally, the AdS minimum and the associated local maximum are separated from the Minkowski vacuum by a singularity at $T = \frac12 ( \frac{3 \xi}{2})^{2/3}$ which originates from a pole in the inverse K\"ahler metric.\footnote{This is related to the fact that the dilaton is assumed to be integrated out.} Thus, the $\alpha'$-correction to the K\"ahler potential allows for a separate local Minkowski vacuum that, contrary to the KKLT scenario, is not an uplifted AdS minimum. 

After this discussion of the vacuum structure produced by K\"ahler Uplifting, we can again couple chaotic inflation and investigate the effective inflaton potential.
%

\subsection{K\"ahler Uplifting and chaotic inflation: analytic approach}

As before, to simplify the discussion we assume that the interactions between modulus and inflaton sector are purely gravitational. Hence, we study the theory defined by 
\begin{subequations}\label{eq:WKKU}\begin{align}
W &= W_0 + A e^{-a T} + \frac12 m \phi^2\,, \\
K &= -2 \ln{\left[ \left(T + \overline T \right)^{3/2} + \xi \right]  + \frac12 \left( \phi + \bar \phi \right)^2}\,.
\end{align}\end{subequations}
Again, since we choose real superpotential parameters only the real part of $T$ is affected by inflation. Hence, we set $T= \overline T$ in the scalar potential. The $\varphi$-dependence of the superpotential leads to a deformation of the Minkowski vacuum and the associated local maximum, which are now determined by the following equation for the covariant derivative,
\begin{align}\begin{split}\label{DTup}
D_T W &= \frac12 Y W\left(1 + \sqrt{1 - \frac{Z}{W^2}m^2 \varphi^2}\right) \\
&= -\frac{3W}{2T} - \frac{3}{8aT^2}\frac{m^2\varphi^2}{W} + \mathcal O (T^{-3})\,,
\end{split}
\end{align}
where the function $Z=\mathcal O (aT)^{-1}$ is defined in Eq.~\eqref{KU4}. Again, this equation implicitly determines $T(\varphi) = T_0 + \delta T(\varphi)$. Since the modulus F-term in this case is bigger than in KKLT, at leading order it cancels the negative contribution to the inflaton potential. At leading order in $\delta T$, $\eta_0$ and $T_0^{-1}$ it is simply
\begin{align}\label{Vupleading}
V = \frac12 \tilde m^2 \varphi^2 + \mathcal{O} (\delta T, \eta_0, T_0^{-1}) \,.
\end{align}
There are two upper bounds on the value of the inflaton field. First, the F-term potential of the inflaton should not exceed the height of the modulus barrier. Second, the expression in Eq.~\eqref{DTup} should yield real values for $D_T W$. Consequently,
\begin{align}\begin{split}\label{eq:KUConstraints}
\tilde m \varphi^2 &\lesssim 4 m_{3/2}\,,\\
\tilde m^2 \varphi^2 &\lesssim \eta_0 m_{3/2}^2 \,.
\end{split}\end{align}

Starting from Eq.~\eqref{DTup} the shift in the modulus field can again be obtained by expanding the covariant derivative in $\delta T$ and $\eta_0$. The leading-order result reads
\begin{align}\label{KUshift}
\frac{\delta T}{T_0} =  \frac{\tilde m^2 \varphi^2}{5 \eta_0 m_{3/2}^2 }
- \frac{9 \tilde m\varphi^2}{20 m_{3/2}} + \dots \,,
\end{align}
where the dots denote higher-order terms in $\eta_0$ and $T_0^{-1}$. Clearly, if the conditions \eqref{eq:KUConstraints} are fulfilled the expansion converges. Expanding the inflaton potential in $\frac{\delta T}{T_0}$ and $\eta_0$, we find at leading order
\begin{align}\label{eq:KUEffPot}
V(\varphi) &\approx \frac12 \tilde m^2 \varphi^2 - \frac{3\eta_0}{4} \tilde m m_{3/2} \varphi^2 - \frac{3}{20 \eta_0} \frac{\tilde m^4 \varphi^4}{m_{3/2}^2} + \frac{27}{40} \frac{\tilde m^3 \varphi^4}{m_{3/2}} - \frac{183 \eta_0}{320} \tilde m^2 \varphi^4 +\dots \,,
\end{align}
which contains negative quartic terms in the inflaton field, analogous to the KKLT case. This time, however, they are suppressed by factors of $\frac{\delta T}{T_0}$ or $\eta_0$.

As in Sec.~\ref{sec:AnalyticKKLT} we can obtain the same result by means of a Taylor expansion of the scalar potential, i.e., by minimizing the expression Eq.~\eqref{eq:KUTaylor}. The modulus shift is inversely proportional to $m_T^2$, and can be written as
\begin{align}\label{KUshift2}
\frac{\delta T}{T_0} =  \frac{4\tilde m^2 \varphi^2 - 9 \eta_0 \tilde m m_{3/2} \varphi^2}{4 m_T^2} + \dots \,.
\end{align}
The first term in the numerator is the leading order inflaton uplift of the potential and the second terms arises due to the incomplete no-scale cancellation at the shifted modulus vacuum expectation value, 
\begin{align}
\delta V \propto K^{T\overline T} |D_T W|^2 - 3|W|^2 \sim \eta |W|^2\,.
\end{align}

The procedure to find the effective potential is significantly simplified by expanding all quantities in powers of $\eta_0$. Since, in this case, $T$ is the only field which contributes to supersymmetry breaking in the vacuum and $m_{3/2}$ is generically very large, the general formula Eq.~\eqref{sp14} does not apply. 
However, it is possible to obtain Eq.~\eqref{eq:KUEffPot} by applying the most general result Eq.~\eqref{eq:GeneralEffPot}, which does not contain assumptions about the scale of supersymmetry breaking. 

In the following we study the phenomenology of inflation resulting from this effective potential in two numerical examples. To this end, it is instructive to rewrite the effective potential as
\begin{align}\label{eq:VKU}
V(\varphi) \approx \frac12 \tilde m^2 \varphi^2 \left( 1 - \frac{3}{10 \eta_0} \frac{\tilde m^2}{m_{3/2}^2} \varphi^2 \right)  - \frac{3 \eta_0}{4} \tilde m m_{3/2} \varphi^2 \left( 1 + \frac{61}{80}\frac{ \tilde m}{m_{3/2}} \varphi^2 \right)\,.
\end{align}
At leading order $V(\varphi)$ consists of two quadratic terms and one relevant correction to each, suppressed by one power of $\frac{H^2}{V_\text B}$. The second piece in Eq.~\eqref{eq:VKU} is very similar to the leading-order potential found in the KKLT case, but is suppressed by one power of $\eta_0$. This means that the supersymmetric mass term for $\varphi$ can drive inflation as well. Before discussing inflation in more detail, let us remark that to guarantee stability of $T$ we require $H^2 < V_\text B$. Using Eq.~\eqref{eq:KUBarrier} this leads to a generic bound on the gravitino mass,
\begin{align}\label{eq:KUGenB}
m_{3/2} > \frac{H}{\sqrt{\eta}} \sim \frac{10^{-4}}{\sqrt{\eta}}\,.
\end{align}

%

\subsection{Numerical examples}
\label{sec:KUNumerical}

Starting from the effective potential Eq.~\eqref{eq:VKU} we can distinguish two cases. Inflation can either be driven by the supersymmetric term proportional to $\tilde m^2 \varphi^2 $, or by the bilinear soft term proportional to $\tilde m m_{3/2} \varphi^2$. 

\begin{figure}[t]
\centering
\includegraphics[width=1\textwidth]{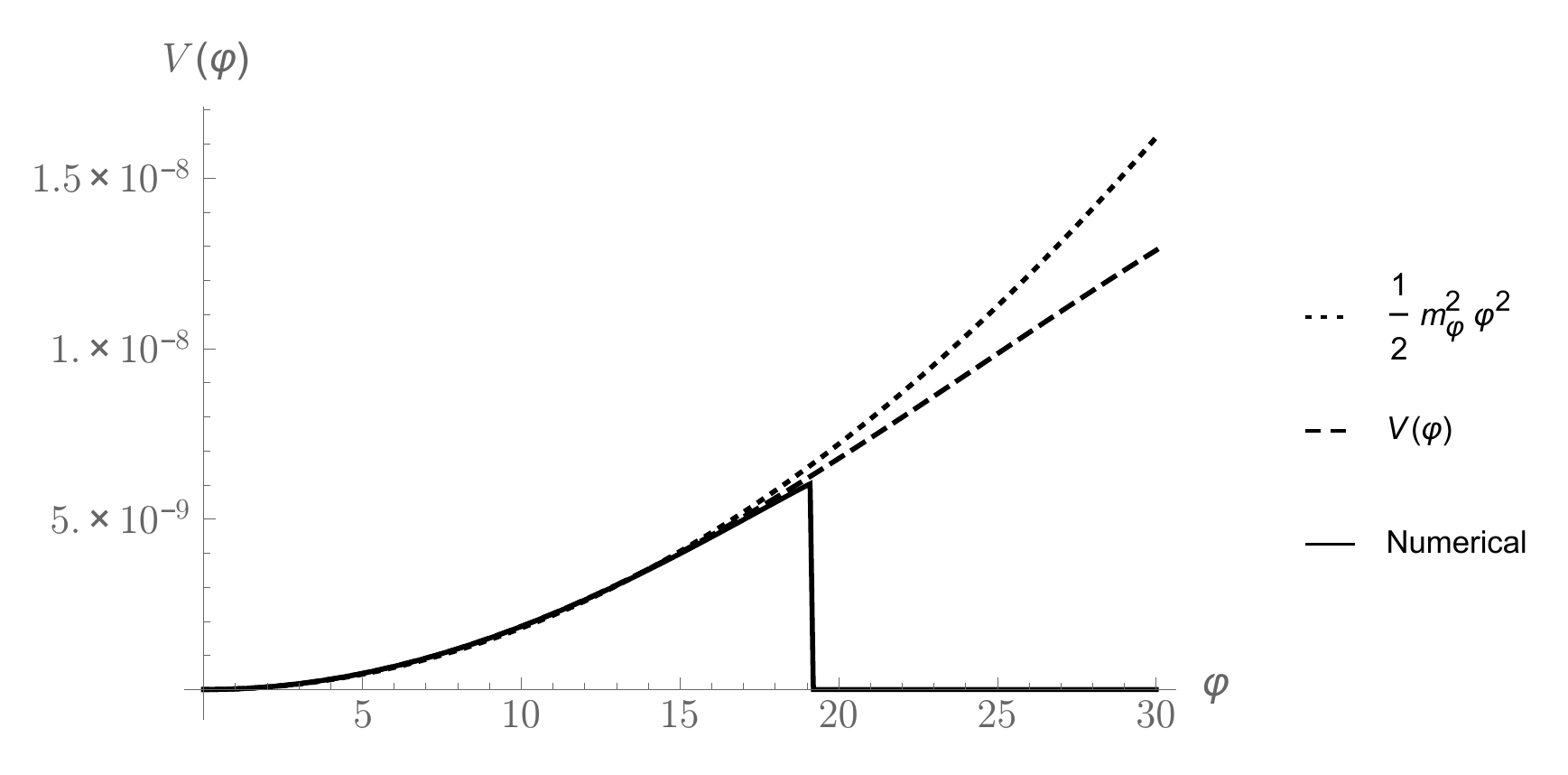} 
\caption{Effective inflaton potential in K\"ahler Uplifting for $W_0 = 4.67$, $A=-3.4 \times 10^{-4}$, $a = \frac{2 \pi}{30}$, $m=8 \times 10^{-4}$, and $\xi = 0.0047$. With these parameters we find $T_0 = 11.9$, $m_{3/2} =0.04$, and $\eta_0 = 2 \times 10^{-5}$. The dotted line denotes a purely quadratic potential with $m_\varphi = 6 \times 10^{-6}$ imposed by COBE normalization. The dashed line is the effective potential Eq.~\eqref{eq:KUEffPot} evaluated at all orders in $\eta$. The solid line is obtained numerically by setting the modulus to its minimum value at each value of $\varphi$. In this case, modulus destabilization occurs at $\varphi_\text c \approx 19$. Again, Eq.~\eqref{eq:KUEffPot} and the dashed line are only meaningful for $\varphi < \varphi_\text c$. \label{fig:PlotKU2}}
\end{figure}

\paragraph{The supersymmetric term dominates}
$\,$ \\
If $\eta_0 m_{3/2} \ll \tilde m$ chaotic inflation may be realized in the ``traditional'' sense. The leading-order potential in this parameter regime is simply the first piece of Eq.~\eqref{eq:VKU}, i.e.,
\begin{align}
V (\varphi) \approx \frac12 \tilde m^2 \varphi^2 \left( 1 - \frac{3}{10 \eta_0} \frac{\tilde m^2}{m_{3/2}^2} \varphi^2 \right)\,.
\end{align}
The viable parameter regime in this scenario is particularly constrained. On the one hand, $\eta_0 m_{3/2}$ must be small for the soft term to be suppressed. On the other hand, $\eta_0 m_{3/2}^2$ must be large enough to guarantee a high barrier in the modulus potential. Specifically, we find
\begin{align}
m_{3/2} \gg \frac{\tilde m^2 \varphi_\star^2}{\eta_0 m_{3/2}} \gg \tilde m \varphi_\star^2 \gtrsim 10 H \sim 10^{-3} \,.
\end{align}
A suitable example is illustrated in Fig.~\ref{fig:PlotKU2}. As expected, the parameter choices are quite elaborate, especially from the perspective of string theory. Specifically, the hierarchy between $W_0$ and $A$ as well as the size of $\eta_0$ are rather particular. With such a small value of $\xi$ it is doubtful whether the string coupling can be small enough to allow for a perturbative description of the theory.

If one ignores this problem inflation can be realized and we find for the solid line 
\begin{align}\begin{split}
n_\text s &= 0.966 \,, \\
r &= 0.116\,,
\end{split}\end{align}
 for $\varphi_\star \approx 15.2$. The modulus is destabilized at $\varphi_\text c \approx 19$.

\begin{figure}[t]
\centering
\includegraphics[width=1\textwidth]{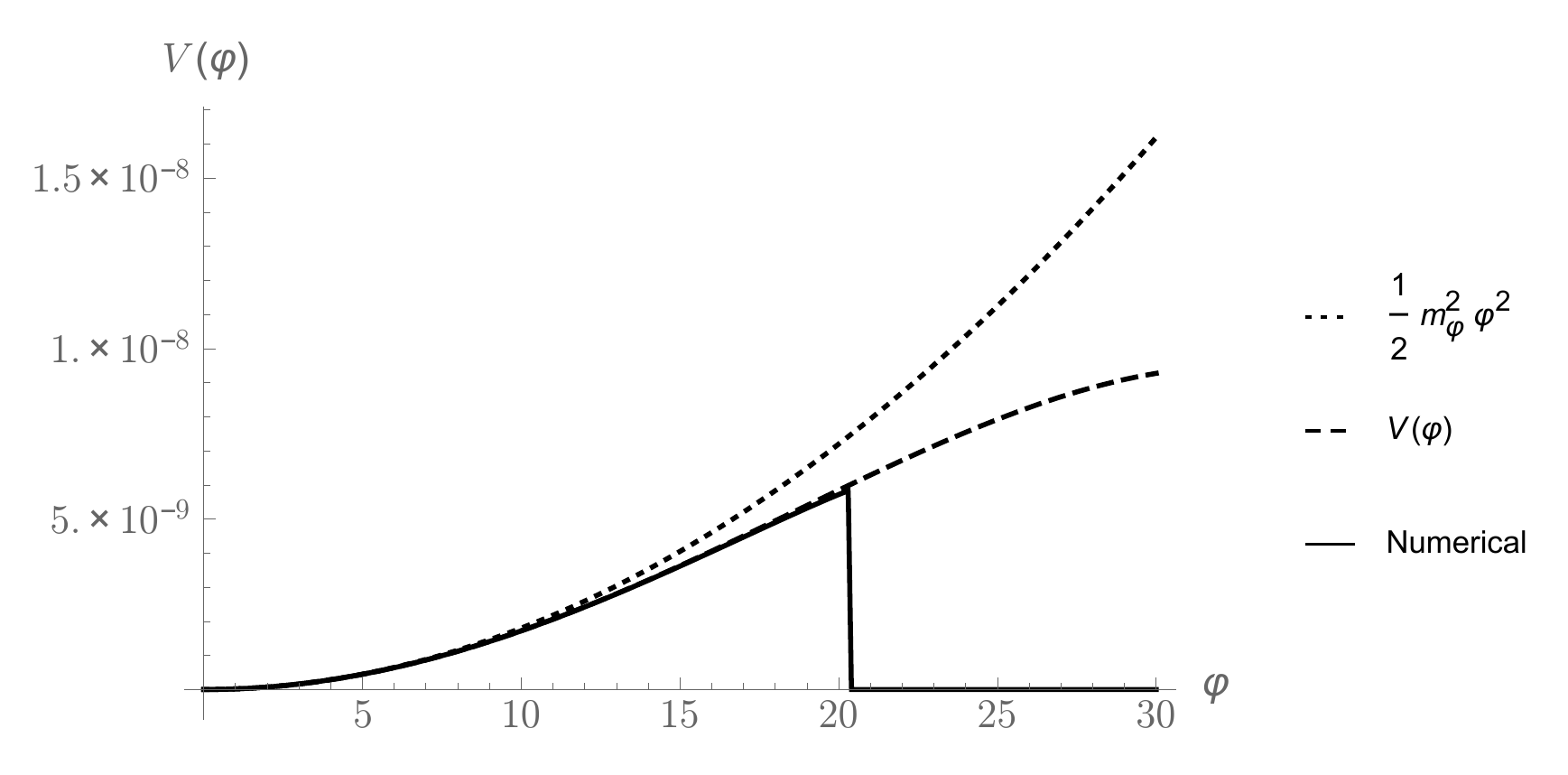} 
\caption{Effective inflaton potential in K\"ahler Uplifting for $W_0 = 0.23$, $A=-0.008$, $a = \frac{2 \pi}{30}$, $m=-1.37 \times 10^{-4}$, and $\xi = 2.29$. With these parameters we find $T_0 = 11.8$, $m_{3/2} =0.002$, and $\eta_0 = 0.01$. The dotted line denotes a purely quadratic potential with $m_\varphi= 6 \times 10^{-6}$ imposed by COBE normalization. The dashed line is the effective potential Eq.~\eqref{eq:KUEffPot} evaluated at all orders in $\eta$. The solid line is obtained numerically by setting the modulus to its minimum value at each value of $\varphi$. In this setup, modulus destabilization occurs at $\varphi_\text c \approx 20$. Again, Eq.~\eqref{eq:KUEffPot} and the dashed line are only meaningful for $\varphi < \varphi_\text c$. \label{fig:PlotKU1}}
\end{figure}

\paragraph{The bilinear soft term dominates}
$\,$ \\
In this respect, the scenario $\eta_0 m_{3/2} \gg \tilde m$ seems slightly more appealing since it can be realized with more realistic choices for the input parameters. The leading-order potential becomes 
\begin{align}
V (\varphi) \approx - \frac{3 \eta_0}{4} \tilde m m_{3/2} \varphi^2 \left( 1 + \frac{61}{80}\frac{ \tilde m}{m_{3/2}} \varphi^2 \right)\,.
\end{align}
Notice the sign difference of the soft term compared to KKLT. Since $\eta_0 > 0$ this means that $\tilde m$ and $m_{3/2}$ must have opposite signs for inflation to work in this parameter regime. COBE normalization imposes $\sqrt{|\eta_0 \tilde m m_{3/2}|} \sim 5 \times 10^{-6}$. Since $\eta_0$ is allowed to be larger in this case, the only bound on $m_{3/2}$ is the generic one, \eqref{eq:KUGenB}. An example is depicted in Fig.~\ref{fig:PlotKU1}.

The corresponding CMB observables are found to be 
\begin{align}\begin{split}
n_\text s &= 0.965\,, \\
r &= 0.107\,,
\end{split}\end{align}
at $\varphi_\star \approx 15$. In this case, the modulus is destabilized at $\varphi_\text c \approx 20$.
%

\section{Chaotic inflation and the Large Volume Scenario}
\label{sec:LVS}

%

\subsection{LVS moduli stabilization and uplift}

Another well-known example of moduli stabilization with spontaneously broken supersymmetry is the Large Volume Scenario developed in \cite{Balasubramanian:2005zx}. It is based on the observation that, for certain types of CY compactifications with multiple K\"ahler moduli, the scalar potential may have a non-supersymmetric AdS minimum at exponentially large volume. A particularly simple example of this type is given by a ``swiss-cheese'' CY manifold with a single ``hole'', i.e., a manifold whose volume is parameterized by 
\begin{align}
\mathcal V = (T_\text b + \overline T_\text b)^{3/2} - (T_\text s + \overline T_\text s)^{3/2}\,,
\end{align}
where $T_\text b$ is the K\"ahler modulus of some big four-cycle, i.e., the ``cheese'', and $T_\text s$ controls the volume of a small four-cycle, the ``hole''. The simplest setup for a Large Volume Scenario is then described by 
\begin{align}
W = W_0 + A e^{- a T_\text s}\,,
\end{align}
and 
\begin{align}
K = -2 \ln (\mathcal V + \xi)\,,
\end{align}
with $\xi$ defined as in Sec.~\ref{sec:KU}. As in the previous examples we consider real superpotential parameters, and hence restrict our attention to the real parts of the moduli, i.e., we set $T_\text{b,s} = \overline T_\text{b,s}$ in the following. 

The extrema of the potential satisfy the two equations $\partial_{T_\text b} V = \partial_{T_\text s} V = 0$. Since the superpotential does not depend on $T_\text b$, they lead to two quadratic equations for $D_{T_\text s}W$ which can be rewritten as
\begin{align}\label{LVSvac}
D_{T_\text s} W= \tilde Y W\,, \qquad \tilde Z_i = 0 \,.
\end{align}
The functions $\tilde Y$ and $\tilde Z_i$ are given in Appendix~\ref{app:KU}. Assuming that $\mathcal V$ is large and expanding $K_{T_\text s}$ and $\tilde Y$ in powers of $\mathcal V^{-1}$, the equation for $D_{T_\text s}W$ yields
\begin{align}\label{LVSAdS}
\left.\partial_{T_s} W\right|_{T_0}  = a A e^{-aT_0} \approx \frac{3(2T_0)^{1/2}}{2\mathcal V_0} W_0 \,.
\end{align}
Remarkably, this equation coincides with Eq.~\eqref{KUAdS} for the AdS minimum in K\"ahler Uplifting after the replacement $\xi \rightarrow \mathcal V_0$, which corresponds to the large volume limit in the LVS scenario. Eq.~\eqref{LVSAdS} determines the volume in terms of $T_0$,
\begin{align}\label{volume}
\mathcal V_0  &\approx \frac{3 \sqrt{T_0} e^{a T_0} W_0}{\sqrt2 a A} \left( 1-\frac{3}{4 a T_0} \right)\,,
\end{align}
at next-to-leading order in $(a T_0)^{-1}$. The second equation in \eqref{LVSvac} determines the value of $T_\text s$. Using the large volume expansions for the functions $\tilde Z_i$, cf.~Eqs.~\eqref{LVS17} and \eqref{LVS18}, we find
\begin{align}\label{Ts}
T_0 \approx \frac{\xi^{2/3}}{2}\left(1 + \frac{2}{3a\xi^{2/3}}\right) + \mathcal O \left((a\xi^{2/3})^{-2}\right)\,.
\end{align}
At leading order in $\mathcal V^{-1}$, $T_0$ only depends on $\xi$ and $a$, as in K\"ahler Uplifting.
Eqs.~\eqref{volume} and \eqref{Ts} can also be obtained by considering the scalar potential in the large volume limit,
\begin{align}\label{eq:LVSPot}
V  \approx \frac{2 \sqrt 2 \, a^2 A^2 \sqrt{T_\text{s}}\, e^{-2 a T_\text{s}} }{3 \mathcal V} - \frac{4 a A W_0  T_\text{s} \,e^{-a T_\text{s}}}{\mathcal V^2} + \frac{3 \xi W_0^2}{2 \mathcal V^3}\,.
\end{align}
To obtain this form the imaginary part of $T_\text s$ has been fixed at $\langle \text{Im}\, T_\text s\rangle = \frac{\pi}{a}$. In this case, $W_0$ and $A$ must have the same sign for the stabilization mechanism to work. Minimizing $V$ with respect to $\mathcal V$ and $T_\text s$ one finds the local AdS minimum with the values $\mathcal V_0$ and $T_0$ given above.

The depth of the AdS vacuum is 
\begin{align}
V_\text{AdS} \sim - \frac{W_0^2}{\mathcal V_0^3}\,,
\end{align}
rather than $W_0^2/\mathcal V_0^2$ as one may naively expect. This is due to the approximate no-scale cancellation between $F_{T_\text b}$ and $W_0^2$. To achieve a complete uplift to a Minkowski vacuum we employ, once more, a Polonyi field $X$ as a toy example. Treating the uplift in the same way as in KKLT moduli stabilization, we assume that $X$ is stabilized with a nearly-vanishing vacuum expectation value.\footnote{Indeed it is possible to verify that, once coupled to chaotic inflation, the displacement of $X$ is again negligible compared to that of $\mathcal V$ and $T_\text s$.} However, in the LVS scheme the quartic term in the K\"ahler potential is not required as $X$ is stabilized by its soft mass term. The contribution of the Polonyi field then amounts to a term $V_\text{up} = \frac{f^2}{\mathcal V^2}$ in the scalar potential. To cancel the cosmological constant in the vacuum, it must be 
\begin{align}\label{chi}
f^2 \approx  \chi_0 W_0^2 \,, \qquad   \chi_0 = \frac{9\sqrt{2T_0}}{2 a \mathcal V_0}\,,
\end{align}
up to terms suppressed by higher powers of $\mathcal V$ or $a T_\text s$.  Here, $\mathcal V_0$ and $T_0$ denote the values of the two real fields in the uplifted vacuum. Note that $\chi_0$ plays a role analogous to the parameter $\eta_0$ in K\"ahler Uplifting. The expression for the volume is still given by
Eq.~\eqref{volume}, where $T_0$ is now the shifted modulus
\begin{align}
T_0 \approx \frac{\xi^{2/3}}{2}\left(1 + \frac{2}{a\xi^{2/3}}\right) + \mathcal O \left((a\xi^{2/3})^{-2}\right)\,
\end{align}
The F-terms of the fields in this vacuum are given by 
\begin{align}
F_{T_\text b} \approx - \sqrt 3 \, \frac{ W_0}{\mathcal V_0} \,,\qquad
F_{T_\text s} \approx \sqrt{6 a T_0\chi_0}\, \frac{W_0}{\mathcal V_0}\,, \qquad
F_X \approx  \sqrt{\chi_0}\, \frac{W_0}{\mathcal V_0} \,.
\end{align}
Clearly, the dominant contribution to supersymmetry breaking comes from the volume mode. As expected, the uplift sector is important to cancel the cosmological constant but its contribution to supersymmetry breaking is suppressed in the large volume limit. The corresponding gravitino mass is, again,
\begin{align}
m_{3/2} \approx \frac{W_0}{\mathcal V_0}\,,
\end{align}
up to terms suppressed by higher powers of the inverse volume or $a T_0$. The masses of the canonically normalized moduli are, schematically\footnote{Note that the axion of $T_\text b$ is exactly massless and thus irrelevant during inflation. The axion of $T_\text s$ is stabilized at the same mass scale as the real part of $T_\text s$.}
\begin{align}\label{eq:LVSMasses}
m_{T_\text b} \sim \frac{W_0}{\mathcal V_0^{3/2}}\,, \qquad m_{T_\text s} \sim \frac{W_0}{\mathcal V_0}\,.
\end{align}
The uplifted vacuum is protected by a potential barrier of height 
\begin{align}\label{eq:LVSBarrier}
V_\text B \sim \frac{m_{3/2}^2}{\mathcal V_0}\,.
\end{align}

Although the structure of this vacuum is more complicated than in the previous two cases, the coupling of chaotic inflation works in the same way. As will become clear in the following, the results are qualitatively similar.

%

\subsection{LVS and chaotic inflation}

Our starting point for the coupled model is this time  
\begin{align}
W &= W_0 + A e^{-a T_\text{s}} + f X + \frac12 m \phi^2\,, \\
K &= -2 \ln{\left[ \left(T_\text{b} + \overline T_\text{b} \right)^{3/2} - \left(T_\text{s} + \overline T_\text{s} \right)^{3/2} + \xi \right] + k\!\left(|X|^2\right) + \frac12 \left( \phi + \bar \phi \right)^2}\,.
\end{align}
The uplift sector is treated as described above, since it is safe to neglect its influence on inflation. The scalar potential at leading order in $\mathcal V^{-1}$ reads
\begin{align}\begin{split}
V &= \frac{2 \sqrt 2 \, a^2 A^2 \sqrt{T_\text{s}}\, e^{-2 a T_\text{s}} }{3 \mathcal V} - \frac{16 a A T_\text{s} \,e^{-a T_\text{s}}\left(4W_0 - m \varphi^2 \right)}{\mathcal V^2} \\ &+ \frac{3 \xi \left(4 W_0 -  m \varphi^2 \right)^2}{32 \mathcal V^3} + \frac{(\mathcal V - 2\xi)\left(f^2+\frac12 m^2 \varphi^2\right)}{\mathcal V^3}\,.
\end{split}\end{align}
Comparing this expression to Eq.~\eqref{eq:LVSPot} we observe that, in principle, the contribution of the inflaton can be absorbed in a redefinition of $W_0$ and $f$. As before, we treat inflation as a perturbation of the true vacuum. Hence, we naively expect chaotic inflation to be successful in LVS as long as 
\begin{align}\label{eq:LVSNaiveConstraints}
m^2 \varphi^2 \ll f^2\,, \qquad m \varphi^2 \ll W_0\,,
\end{align}
neglecting order-one coefficients. It will become clear in the following that these two conditions precisely guarantee that the inflaton energy density does not destabilize the moduli.

To compute the effective inflaton potential we have to take the displacements of both moduli into account. Hence, we expand the potential around 
\begin{align}
\delta \mathcal V = \mathcal V - \mathcal V_0\,, \qquad \delta T_\text s = T_\text s - T_0\,.
\end{align}
Minimizing the result with respect to both shifts yields
\begin{subequations}\begin{align}
\frac{\delta \mathcal V}{\mathcal V_0} &\approx  \frac{\tilde m^2 \varphi^2}{\chi_0 m_{3/2}^2} + \frac{\tilde m \varphi^2}{4 m_{3/2}}\,, \\
\frac{\delta T_\text s}{T_0} &\approx \frac{\tilde m^2 \varphi^2}{aT_0 \chi_0 m_{3/2}^2} + \frac{\tilde m \varphi^2}{2aT_0 m_{3/2}}\,,
\end{align}\end{subequations}
up to terms suppressed by higher powers of $\mathcal V^{-1}$ or $(a T_0)^{-1}$. Note that the shifts have the same form as in K\"ahler Uplifting, cf.~Eq.~\eqref{KUshift}. Furthermore, the displacement of $T_\text s$ is relatively suppressed by one power of $\mathcal V_0$. This is to be expected because $T_\text s$ is the heavier of the two moduli. Nonetheless, $\delta T_\text s$ must be taken into account to find the correct leading-order result.

Integrating out the displacements of both moduli, we are left with the leading-order effective potential
\begin{align}\begin{split}\label{eq:LVSEffPot}
V (\varphi) \approx \frac12 \tilde m^2 \varphi^2 + \frac{\chi_0}{4} \tilde m m_{3/2}\varphi^2 -
\frac{1}{2\chi_0}\frac{\tilde m^4 \varphi^4}{m_{3/2}^2} - \frac14 \frac{\tilde m^3 \varphi^4}{m_{3/2}} -
\frac{\chi_0}{16 aT_0} \tilde m^2 \varphi^4 \,.
\end{split}\end{align}
We refrain from rewriting this unwieldy expression in terms of the moduli masses, but the idea is the same as in our previous examples. Some of the correction terms are suppressed by inverse powers of $m_{T_\text b}$ and $m_{T_\text s}$ and vanish in the limit of very heavy moduli. Others, like the supersymmetry breaking second term in Eq.~\eqref{eq:LVSEffPot} grow with the moduli masses, and hence do not vanish. As in the previous examples, the region where $V(\varphi)$ is unbounded from below is never reached since the moduli are destabilized at smaller values of $\varphi$.

As in our model with K\"ahler Uplifting we rewrite the effective potential to study inflation. In particular,
\begin{align}\label{eq:VeffLVS}
V(\varphi) \approx \frac12 \tilde m^2 \varphi^2 \left( 1 -\frac{1}{\chi_0}\frac{\tilde m^2}{m_{3/2}^2} \varphi^2 \right) + \frac{\chi_0}{4} \tilde m m_{3/2}\varphi^2 \left(1 -\frac{1}{4 a T_0} \frac{\tilde m}{m_{3/2}} \varphi^2 \right) \,.
\end{align}
Again, $V(\varphi)$ contains a supersymmetric mass term and a bilinear soft term -- suppressed by one power of $\chi_0$ --, both with a correction proportional to $\frac{H^2}{V_\text B}$. By requiring the barrier to be larger than the Hubble scale during inflation, the gravitino mass is generically constrained as follows,
\begin{align}\label{eq:LVSGenB}
m_{3/2} > H \sqrt{\mathcal V_0} \sim 10^{-4} \sqrt{\mathcal V_0}\,.
\end{align}
As before, this constraint is equivalent to demanding that $\varphi$ is not large enough to uplift the modulus minimum to a saddle point. 

%

\subsection{Numerical examples}

Based on the effective potential Eq.~\eqref{eq:VeffLVS} we can distinguish two cases in which 60 e-folds of inflation may be realized. 

\paragraph{The supersymmetric term dominates}
$\,$ \\
If $\tilde m \gg \chi_0 m_{3/2} \sim m_{3/2}/\mathcal V_0$, in principle the supersymmetric quadratic term in Eq.~\eqref{eq:VeffLVS} could dominate, yielding the leading-order potential
\begin{align}\label{eq:LVS21}
V (\varphi) &\approx \frac12 \tilde m^2 \varphi^2 \left( 1 -\frac{1}{\chi_0}\frac{\tilde m^2}{m_{3/2}^2} \varphi^2 \right) \,.
\end{align}
However, this scenario is excluded by a consistency requirement of the LVS scheme. Specifically, the gravitino mass must not exceed the Kaluza-Klein scale which, as discussed in \cite{Cicoli:2013swa}, means that $W_0 \ll \mathcal V_0^{1/3}$. Requiring the supersymmetric term to be larger than the soft term while both moduli are stabilized always violates this bound. For different effects related to the Kaluza-Klein scale, cf.~\cite{Chialva:2014rla,Talaganis:2014ida}.

\begin{figure}[t]
\centering
\includegraphics[width=1\textwidth]{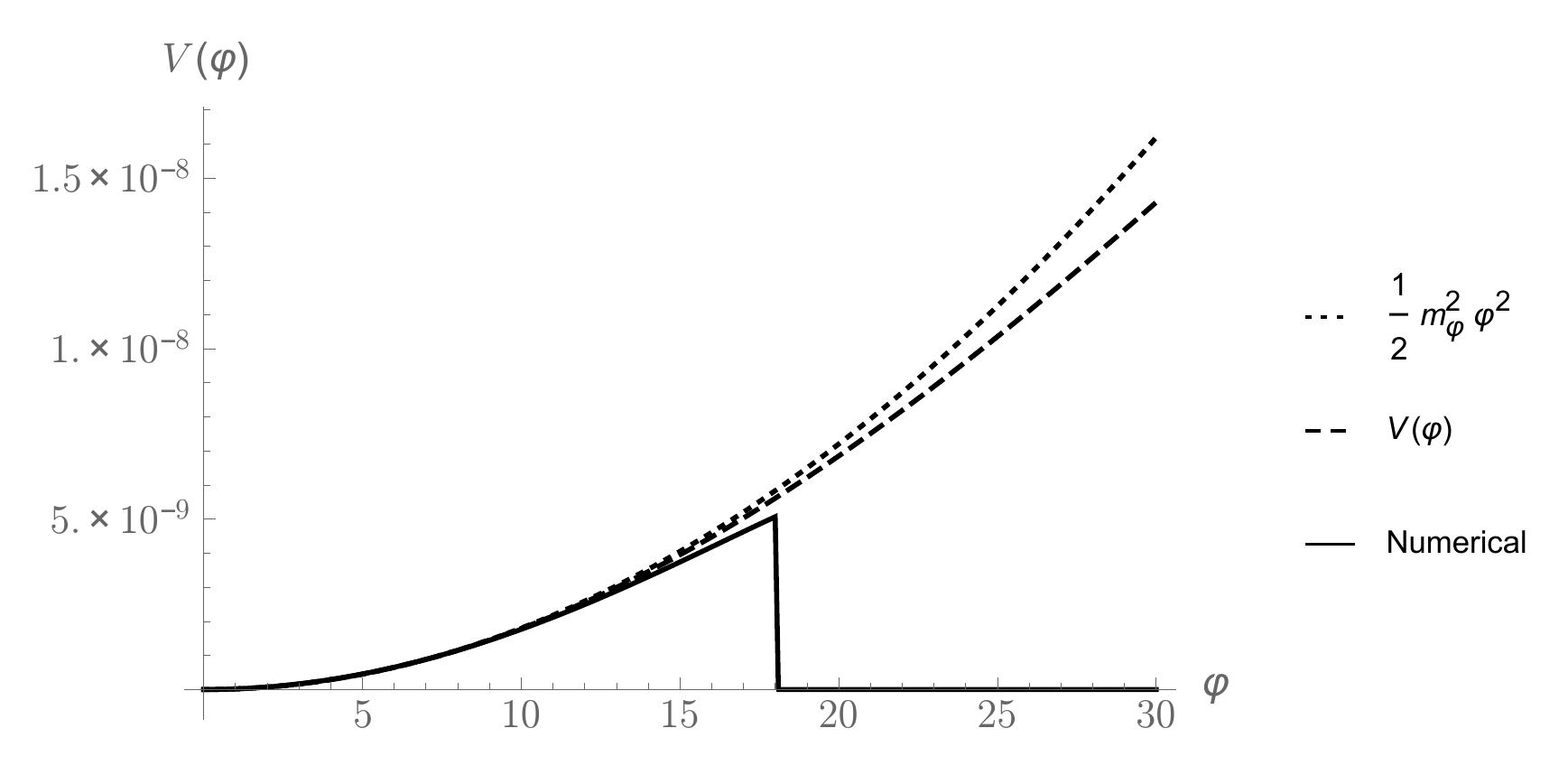} 
\caption{Effective inflaton potential in LVS for $W_0 = 1$, $A=0.13$, $a = 2 \pi$, $m=5.8 \times 10^{-4}$, and $\xi = 1.25$. With these parameters we find $T_0 = 0.75$, $\mathcal V_0 = 200$, and $m_{3/2} =0.005$. The dotted line denotes a purely quadratic potential with $m_\varphi= 6 \times 10^{-6}$ imposed by COBE normalization. The dashed line is the effective potential Eq.~\eqref{eq:LVSEffPot} evaluated at all orders in $a T_0$. The solid line is obtained numerically by setting the modulus to its minimum value at each value of $\varphi$. Since the barrier height and Hubble scale are the same as in the previous example, modulus destabilization occurs at $\varphi_\text c \approx 18$. Again, Eq.~\eqref{eq:LVSEffPot} and the dashed line are only meaningful for $\varphi < \varphi_\text c$. Notice that the difference between the dashed and the solid line is comparably large in this example. This is because the relatively small value of $\mathcal V_0$ limits the precision of the expansion in $\mathcal V^{-1}$.  \label{fig:PlotLVS1}}
\end{figure}

\paragraph{The bilinear soft term dominates}
$\,$ \\
If, on the other hand, $\tilde m \ll \chi_0 m_{3/2} \sim m_{3/2}/\mathcal V_0$, the term proportional to $\tilde m m_{3/2}$ may drive inflation. In this case, the leading-order inflaton potential reads
\begin{align}
V (\varphi) &\approx \frac{\chi_0}{4} \tilde m m_{3/2}\varphi^2 \left(1 -\frac{1}{4 a T_0} \frac{\tilde m}{m_{3/2}} \varphi^2 \right) \,.
\end{align}
The gravitino mass is constrained by the generic requirement \eqref{eq:LVSGenB}. Interestingly, by requiring $m_{3/2} < m_\text{KK}$ for consistency, the volume of the compactification manifold is bounded from above, $\mathcal V_0 \lesssim 10^3$. A numerical example for this scenario is depicted in Fig.~\ref{fig:PlotLVS1}.
 
The CMB observables in our example are found to be 
\begin{align}\begin{split}
n_\text s &= 0.964\,, \\
r &= 0.116\,,
\end{split}\end{align}
at $\varphi_\star \approx 15.2$. Modulus destabilization towards the run-away minimum occurs at $\varphi_\text c \approx 18$.

%

\section{Universality and CMB observables}

Let us consider the effective single-field inflaton potential arising in all three example models as well as in the general discussion of Sec.~2. We observe that a simple expression captures all models and their flattening of the inflaton potential by moduli backreaction,
\begin{align}
V(\varphi)=\frac12 m_\varphi^2 \,\varphi^2-\frac14 \lambda \varphi^4\,,\qquad \lambda>0\,.
\end{align}
This expression is valid at leading order in the modulus shift, and thus holds for a certain range $\varphi<\varphi_\text c$ until the moduli are destabilized. 

Due to the negative quartic term the potential has a local maximum at $\varphi_\text{M}=m/\sqrt\lambda$. All three scenarios share the property that the moduli destabilization point occurs to the left of the maximum of the leading-order inflaton potential, 
\begin{align}
\varphi_\text c < \varphi_\text{M}\,.
\end{align}
Hence, $V(\varphi)$ is a good approximation for $\varphi<\varphi_c$. Two parameters determine the effective potential, $m/\sqrt\lambda$ gives the position of the maximum and $m$ fixes the overall normalization of $V(\varphi)$. Thus, we can write the potential in terms of $m$ and $\varphi_\text{M}$,
\begin{align}
V(\varphi)=\frac12 m_\varphi^2\,\varphi^2\,\left(1-\frac{\varphi^2}{2\varphi_\text{M}^2}\right)\,.
\end{align}
As long as $\varphi_\text{M}, \varphi_\text c \gg 1$ inflation can occur to the left of the local maximum. For ${\varphi_\text{M}\to\infty}$ the potential asymptotes to the pure quadratic form. In this limit, the field value $\varphi_\star$ corresponding to $N_e(\varphi_\star)$ e-folds of slow-roll before the end of inflation takes the limiting value $\varphi_\star=2\sqrt{N_e}$, which for $N_e=50-60$ is about $15$.

For decreasing $\varphi_\text{M}$ the 60 e-fold point lies increasingly close to the local maximum and the destabilization point. Thus, for $\varphi_\text{c}\to\varphi_\star$ the inflationary dynamics changes continuously from the quadratic large-field behaviour to a nearly hill-top small-field model. Correspondingly, the scalar spectral index and $r$ are decreased compared to pure quadratic inflation.

Inflaton potentials of this type arise in the context of non-minimally coupled quadratic inflation~\cite{Linde:2011nh} and more recently in subcritical models of D-term hybrid inflation~\cite{Buchmuller:2014rfa,Wieck:2014xxa,Buchmuller:2014dda}. As the leading-order scalar potential is the same for all our models, the CMB observables agree as well. Reproducing the particularly simple form given in~\cite{Buchmuller:2014rfa,Buchmuller:2014dda} one finds
\begin{align}
\epsilon=\frac{2}{\varphi^2}\left(\frac{1-\frac{\varphi^2}{\varphi_\text{M}^2}}{1-\frac{\varphi^2}{2\varphi_\text{M}^2}}\right)^2\,, \qquad
\quad \eta=\frac{2}{\varphi^2}\left(\frac{1-\frac{3\varphi^2}{\varphi_\text{M}^2}}{1-\frac{\varphi^2}{2\varphi_\text{M}^2}}\right)\,.
\end{align}
Extracting $\varphi_\star$ from $N_e=\int_{\varphi_e}^{\varphi_\star} {\rm d}\varphi/\sqrt{2\epsilon}$ we obtain
\begin{align}
\varphi_\star^2=4N_e+2-\frac{4 N_e^2}{\varphi_\text{M}^2}-\frac{8N_e^3}{3\varphi_\text{M}^4}+\frac{4 N_e^4}{3\varphi_\text M^6}+\ldots
\end{align}
where we have used the leading-order expression for the end-point of slow-roll inflation, ${\varphi_\text e=\sqrt 2- \mathcal O (\varphi_\text{M}^{-2})}$. From this it is evident that all our models approach the quadratic inflation limit as $\varphi_\text M \to \infty$.

\begin{figure}[t!]
\centering
\includegraphics[width=1\textwidth]{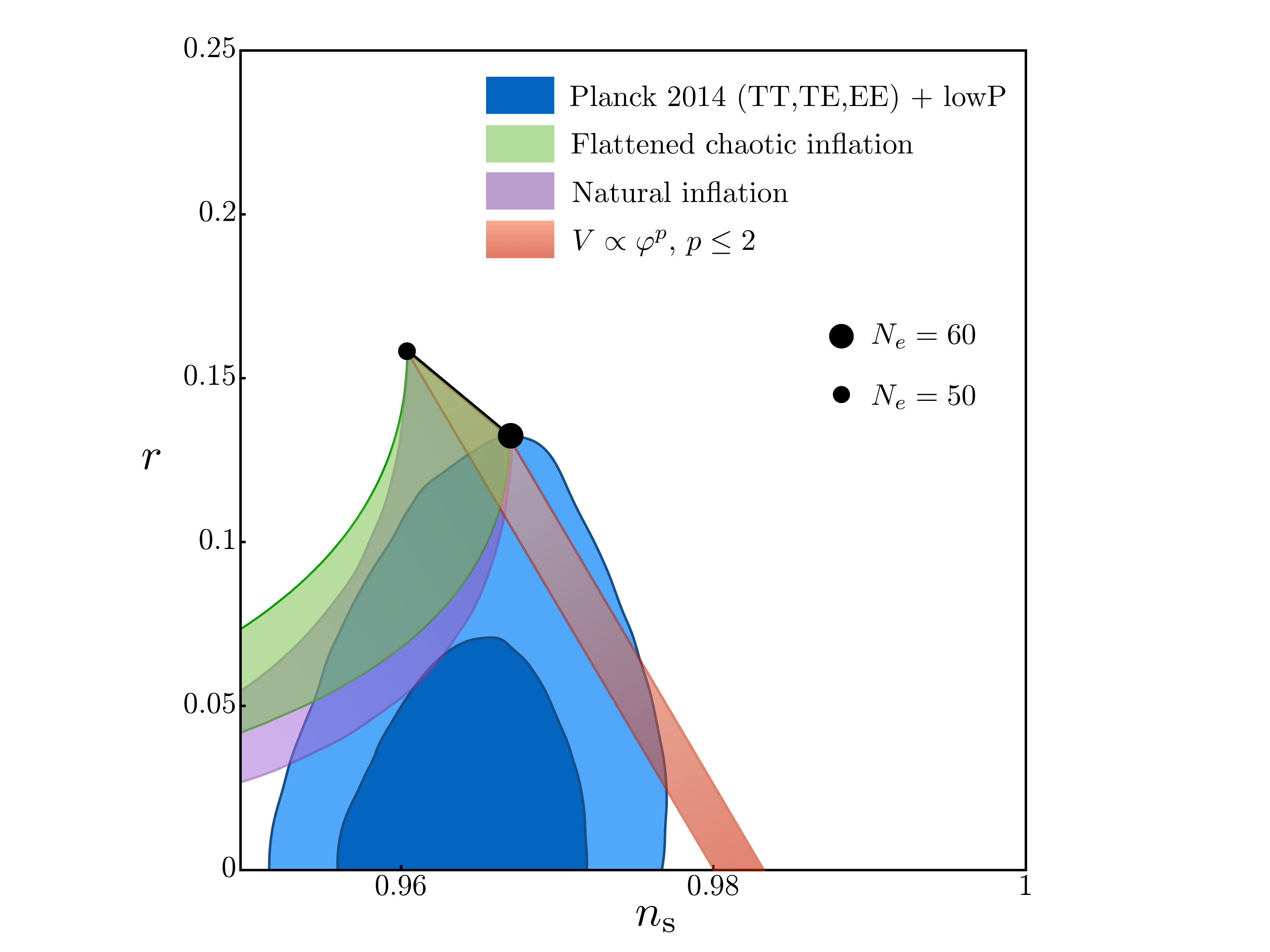} 
\caption{Prediction for the CMB observables $n_\text s$ and $r$ of the leading-order effective inflaton potential. In the limit $\varphi_\text{M}\to \infty$ the observables asymptote to the predictions of pure quadratic inflation. Decreasing $\varphi_\text{M}$ brings the potential increasingly into the hill-top regime. This leads to the green band of decreasing $n_\text s$ and $r$ values spanned by the 60 and 50 e-fold curves. Note, once more, that the regime of true hill-top inflation can actually never be reached because moduli destabilization occurs to the left of the would-be local maximum in $V(\varphi)$ at $\varphi_\text M$. \label{fig:ns-r}}
\end{figure}

Comparison with the full numerical solution for $\varphi_\star$ reveals that the analytic expression above must be given up to ${\cal O}(\varphi_\text{M}^{-6})$ for sufficient accuracy. The terms with inverse powers of $\varphi_\text M$ are given at leading order in $N_e$ to allow for a compact expression. We find that for $n_\text s \gtrsim 0.94$ this form approximates the ensuing values of $n_\text s$ and $r$ to 5\% numerical accuracy compared to the exact coefficients given in \cite{Buchmuller:2014dda}. Plugging back $\varphi_\star$ into the expressions for $\epsilon$ and $\eta$ we can compute the spectral parameters of the curvature and tensor perturbation power spectra
\begin{align}\begin{split}
n_\text s&= 1-6\epsilon(\varphi_\star)+2\eta(\varphi_\star)\,, \\
r&= 16\epsilon(\varphi_\star)\,.
\end{split}\end{align}
at horizon exit. Doing this numerically and comparing the result with the Planck data results in the green band in Fig.~\ref{fig:ns-r} which is identical to the corresponding graph in~\cite{Buchmuller:2014dda}. 
Imposing the constraints on $n_\text s$ and $r$ we find a lower bound on the tensor-to-scalar ratio, $r \gtrsim 0.05$, for $N_e=60$.

Finally, we make an interesting observation. On the one hand, our effective inflation potential arises for all three models studied here as an approximation of, for example, type IIB string theory constructions of axion monodromy inflation with an F-term supergravity description~\cite{Palti:2014kza,Marchesano:2014mla,Hebecker:2014eua,Grimm:2014vva,Ibanez:2014kia,Blumenhagen:2014nba,Hebecker:2014kva,Ibanez:2014swa,Garcia-Etxebarria:2014wla} as well as of models of D-term hybrid inflation~\cite{Buchmuller:2014rfa,Wieck:2014xxa,Buchmuller:2014dda}. Moreover, we found in this work that all our models show a form of polynomial flattening of the naive quadratic inflation potential by subtracting (at leading order) a higher-power monomial term in $\varphi$
\begin{align}
V(\varphi)\sim\varphi^{p_0} f(\varphi)\,,\qquad f(\varphi)=1-c\varphi^2+\ldots \,,\qquad p_0=2\,.
\end{align}
The flattening occurs in a regime with $c\ll 1$ and small higher-oder coefficients. 

On the other hand, there is a large class of models of axion monodromy inflation which feature a form of monomial flattening~\cite{Dong:2010in,Kaloper:2011jz,Dubovsky:2011tu,Kaloper:2014zba,McAllister:2014mpa}. Some of these setups work without a supergravity embedding or with inflation from a sector with non-linearly realized supersymmetry arising from non-supersymmetric compactifications like Riemann surfaces~\cite{Silverstein:2008sg,McAllister:2008hb,Dong:2010in,Dodelson:2013iba,McAllister:2014mpa} while another one involves F-term monodromy on D-branes~\cite{Ibanez:2014swa}. In these constructions a quadratic or quartic inflation potential flattens by suppressing the monomial power $p < p_0$,
 \begin{align}
V(\varphi)\sim\varphi^{p_0} f(\varphi)\,,\qquad f(\varphi)=\varphi^{-\Delta p}\,\qquad p_0 = 2,4\,.
\end{align}
The correlation between the two types of flattening -- polynomial and monomial -- may be due to the different mechanisms of volume stabilization (non-perturbative versus perturbative). In particular, polynomial flattening seems to correlate with models showing spontaneous bulk F-term supersymmetry breaking and non-perturbative volume stabilization (implying CY compactification). We may speculate here that both of these correlations hold more widely.

Moreover, the two types of flattening have quite different observational predictions, with polynomial flattening corresponding to the green band and monomial flattening yielding the red band in Fig.~\ref{fig:ns-r}. Future CMB data may enable us to discriminate between the two types of flattening -- and hence maybe even between classes of string compactifications.

%

\section{Discussion and conclusion}

The aim of this paper is to study the interplay between K\"ahler moduli stabilization and large-field inflation in the context of string-effective supergravity models. We find that if moduli stabilization breaks supersymmetry, the modulus sector never decouples from inflation. On the one hand, supersymmetry breaking induces a bilinear soft mass term for the inflaton which can potentially drive 60 e-folds of slow-roll inflation. On the other hand, the potential contains dangerous terms which destabilize the moduli if the inflaton field exceeds a critical value. 

We have illustrated our results in three prominent models of moduli stabilization: KKLT, K\"ahler Uplifting and the simplest Large Volume Scenario. In all three models we have analyzed corrections to the inflaton potential from supersymmetry breaking and from integrating out the moduli. Although the dominant source of supersymmetry breaking and the structure of vacua differ in the three models, they share a number of common features. First, we find that all of them give rise to an effective inflaton potential of the form
\begin{align}\nonumber
V(\varphi) = \frac12 m_\varphi^2 \, \varphi^2 - \frac14 \lambda \varphi^4\,,
\end{align}
after the moduli have been integrated out. Hence, they share universal predictions for the CMB observables, in particular $r \gtrsim 0.05$. Second, in all models the stability of moduli during inflation imposes a severe lower bound on the scale of supersymmetry breaking. In KKLT this is the well-known bound $m_{3/2} > H$. In K\"ahler Uplifting and the Large Volume Scenario, the moduli masses and the potential barrier are suppressed compared to $m_{3/2}$ due to an approximate no-scale symmetry. This leads to the more stringent constraint $m_{3/2} > H \sqrt{\mathcal{V}}$, where $\mathcal{V}$ denotes the volume of the compactification manifold. Unfortunately, this implies that supersymmetry can no longer protect the flatness of the inflaton potential. This is opposite to chaotic inflation with a stabilizer field, where the gravitino mass must be parametrically smaller than the inflaton mass. Third, in all considered schemes the parameter choices required by successful inflation appear unnatural from the perspective of string theory. Although our analysis is limited to specific examples we believe that this problem is characteristic for a wide class of large-field inflation models coupled to a modulus sector.

Another important caveat is that the initial conditions of inflation must be chosen very carefully. The moduli are destabilized if the energy density of the universe exceeds the barrier protecting their local minimum. In this case, the desired regime of slow-roll inflation is never reached. In this sense, the effective inflation models obtained after integrating out the moduli are no longer ``chaotic''.

%

\subsection*{Acknowledgments}
The authors would like to thank Arthur Hebecker, Koji Ishiwata, Renata Kallosh, Andrei Linde, and Lukas Witkowski for stimulating discussions. This work has been supported by the German Science Foundation (DFG) within the Collaborative Research Center 676 ``Particles, Strings and the Early Universe'' and the SFB-Transregio TR33 ``The Dark Universe'', and by the ERC advanced grant ``MassTeV". C.W. acknowledges support by the Joachim Herz Foundation and would like to thank the CPhT at Ecole Polytechnique for hospitality. E.D. and L.H. thank the Alexander von Humboldt foundation and DESY Hamburg for support and hospitality.

%
\begin{appendix}

%
\section{Moduli masses}
\label{app:Masses}

\subsection{Supergravity mass formulae}

Scalar masses in supergravity with zero cosmological constant are given by \cite{soni}
\begin{align}\begin{split}
m^2_{\alpha \bar \beta} &= e^G \left(G_{\alpha \bar \beta} - R_{\alpha {\bar \beta} \gamma
{\bar \delta}} G^{\gamma} G^{\bar \delta} + \nabla_{\alpha} G^{\bar \gamma} 
 \nabla_{\bar \beta} G_{\bar \gamma} \right) \,, \\
m^2_{\alpha \beta}& = e^G \left(2 \nabla_{\alpha} G_{\beta} + G^{\gamma} \nabla_{\alpha} \nabla_{\beta} G_{\gamma} \right) \,,  \label{m1}
\end{split}\end{align}
without taking canonical normalization into account. Here, $R_{\alpha {\bar \beta} \gamma {\bar \delta}}$ is the Riemann curvature of the K\"ahler manifold and $\Gamma^{\alpha}_{\beta \gamma} = G^{\alpha \bar \alpha}\partial_\beta G_{\gamma \bar \alpha}$. Notice that these expressions can be used to compute physical masses in the ground state of the theory, but not during inflation. 

The fermionic mass matrix, on the other hand, is given by
\begin{align}
({\tilde m_F})_{\alpha \beta} = e^{G/2} (\nabla_{\alpha} G_{\beta} + G_{\alpha} G_{\beta})
\ .  \label{m2}
\end{align}
After extracting the goldstino-gravitino mass mixing, the fermionic mass matrix becomes
\begin{align}
({m_F})_{\alpha \beta} = e^{G/2} \left(\nabla_{\alpha} G_{\beta} + \frac{1}{3} G_{\alpha} G_{\beta}\right) = e^{K/2} \left(D_{\alpha} D_{\beta} W - \frac{2}{3 W} D_{\alpha} W D_{\beta} W \right) \,.  \label{m3}
\end{align}
The fermionic masses also define the supersymmetric contribution to the scalar masses. Hence, we can define the soft scalar mass matrix $m_0$ by subtracting the fermionic mass contribution,
\begin{align}\begin{split}
m^2_{\alpha \bar \beta} &=  ({m_F m_F^{\dagger}})_{\alpha \bar \beta} +
e^G \left(G_{\alpha \bar \beta} - R_{\alpha {\bar \beta} \gamma
{\bar \delta}} G^{\gamma} G^{\bar \delta} + \frac{1}{3} G_{\alpha} G_{\bar \beta} \right) 
\equiv \left({m_F m_F^{\dagger}}\right)_{\alpha \bar \beta} + \left({m_0^2}\right)_{\alpha \bar \beta}\ ,  \\
m^2_{\alpha \beta} &= 2 e^{G/2} ({m_F})_{\alpha \beta} +
 e^G \left(- \frac{2}{3} G_{\alpha} G_{\beta} + G^{\gamma} \nabla_{\alpha} \nabla_{\beta} G_{\gamma} \right) \,,  \label{m4}
\end{split}\end{align}
where $\left({m_F m_F^{\dagger}}\right)_{\alpha \bar \beta} = G^{\gamma \bar \gamma}
({m_F})_{\alpha \gamma} (\overline{m}_F)_{\bar \beta \bar \gamma} \equiv \left(m_S^2\right)_{\alpha \bar \beta}$. Furthermore, it is useful to define the inverse supersymmetric mass matrix,
\begin{align}
\left(m_S^{-2}\right)^{\bar \alpha \delta} = G_{\beta \bar \gamma} \left({m_F}^{-1}\right)^{\beta \delta}
\left(\overline{m}_F^{-1}\right)^{\bar \alpha \bar \gamma}  \,,  \label{m5}
\end{align}
which satisfies the relations
\begin{align}
\left({m_F}\right)_{\alpha \beta} \left(m_S^{-2}\right)^{\alpha \bar \beta}
\ = \ G_{\beta \bar \gamma} (\overline{m}_F^{-1})^{\bar \beta \bar \gamma}
\,, \qquad \left(m_S^{-2}\right)^{\alpha \bar \beta} 
\left(\overline{m}_F\right)_{\bar \beta \bar \gamma} = G_{\beta \bar \gamma} \left({m_F}^{-1}\right)^{\beta \alpha} \,.  \label{m6}
\end{align}


\subsection{Nearly-supersymmetric stabilization}
\label{app:Massapprox}

If the supersymmetric masses are much larger than the supersymmetry breaking scale, ${m_F \gg m_{3/2}}$, we can expand the inverse mass matrix,
\begin{align}
m^2_{\alpha \bar \beta} &= \left(m_S^2\right)_{\alpha \bar \gamma} \left[ \delta_{\bar \beta}^{\bar \gamma} + \left(m_S^{-2}\right)^{\bar \gamma \delta}  \left(m_0^2\right)_{\delta \bar \beta} \right] \nonumber \\
\Rightarrow \quad \left(m^{-2}\right)^{\bar \alpha \beta} &\approx \left(m_S^{-2}\right)^{\bar \beta \beta}  \left[ \delta_{\bar \beta}^{\bar \alpha} - \left(m_S^{-2}\right)^{\bar \alpha \delta}  \left(m_0^2\right)_{\delta \bar \beta} \right] \,.  \label{m7} 
\end{align}
In this limit the holomorphic terms $m^2_{\alpha \beta}$ are small, so that for the inverse of the mass matrix
\begin{align}
{\cal M}^2 = \begin{pmatrix}
m^2_{\alpha {\bar \beta}} & m^2_{\alpha \beta} \\
m^2_{{\bar \alpha} {\bar \beta}} & m^2_{{\bar \alpha} \beta} 
\end{pmatrix} \,,
\end{align}
we find
\begin{align}
{\cal M}^{-2} \approx
\begin{pmatrix}
(m^{-2})^{{\bar \beta} \gamma} & - (m^{-2})^{{\bar \beta} \gamma} m^2_{\gamma \beta} (m^{-2})^{\beta \bar \gamma} \\
- (m^{-2})^{\beta {\bar \alpha}} m^2_{\bar \alpha \bar \beta} (m^{-2})^{\bar \beta \gamma} &  (m^{-2})^{\beta \bar \gamma}
\end{pmatrix} \,.  \label{m8}
\end{align}

%
\section{Details of integrating out supersymmetry-breaking moduli}
\label{app:GeneralExpansion}

\subsection{Obtaining the general result}
\label{app:GeneralExpansion1}

The coefficients of the Taylor series in Eq.~\eqref{sp6} are given by
\begin{subequations}\begin{align}
V_0  &= e^{K_0}  \left\{  K_0^{\alpha, \bar \beta} \left[ W_{\alpha} +K_{0, \alpha } W_\text{mod}  \right] \left[ \overline{{W}}_{\bar \beta} +K_{0, {\bar \beta} }  \overline{W}_\text{mod}  \right] - 3 |W_\text{mod}|^2 \right\} \,, \\
V_1 &= e^{K_0}  \left\{ - \frac{1}{2} K_0^{\alpha \bar \beta} \left( K_{0, {\bar \beta}} D_{\alpha} W_\text{mod} + K_{0, \alpha } \overline{D}_{\bar \beta} \overline{W}_\text{mod} \right) + m K_1^{-1} + \frac{3}{2} (W_\text{mod} + \overline{W}_\text{mod}) \right\} \,, \\
 V_2  &= \frac{1}{4} e^{K_0}  \left\{  K_0^{\alpha \bar \beta} K_{0, \alpha }
K_{0, {\bar \beta} } - 3 \right\}  \,, \label{sp7}
\end{align}\end{subequations}
where $D_{\alpha} W_\text{mod} = W_{\text{mod}, \alpha} + K_{0, \alpha } W_\text{mod}$. Expanding these coefficients at leading order in $\delta T_{\alpha} \ll T_{\alpha,0}$ leads to
\begin{subequations}\begin{align}\label{eqs1}
V_0 (T_{\alpha}, {\overline T}_{\bar\alpha} ) &= \frac{1}{2}  \begin{pmatrix} \delta T_{\alpha} & \delta {\overline T}_{\bar \alpha} \end{pmatrix}  \begin{pmatrix} m^2_{\alpha {\bar \beta}} & m^2_{\alpha \beta} \\ m^2_{{\bar \alpha} {\bar \beta}} & m^2_{{\bar \alpha} \beta}  \end{pmatrix} \begin{pmatrix} \delta {\overline T}_{\bar \beta} \\ \delta {T}_{\beta} \end{pmatrix} + \dots \,, \\
V_1 (T_{\alpha}, {\overline T}_{ \bar \alpha} ) &= V_1 (T_{\alpha,0}, {\overline T}_{\bar \alpha,0} ) + \frac{\partial V_1}{\partial T_{\alpha}} \delta T_{\alpha} + \frac{\partial V_1}{\partial {\overline T}_{\bar \alpha}} \delta {\overline T}_{\bar \alpha} + \dots \,, \\
 V_2 (T_{\alpha}, {\overline T}_{\alpha} ) &=  V_2 (T_{\alpha,0}, {\overline T}_{\bar \alpha,0} )
 + \dots \,, \label{sp9}
\end{align}\end{subequations} 
keeping only the leading-order terms up to fourth order in $\varphi$. $m^2_{\alpha \bar \beta}$ and $m^2_{\alpha \beta}$ denote the mass matrices of the moduli fields in the true vacuum. They can be found in Appendix~\ref{app:Masses}. In the expansion of $V_0$ we have used that the cosmological constant vanishes in the vacuum and that the moduli trace their minima adiabatically. In particular, 
\begin{align}
V_0 (T_{\alpha,0}, {\overline T}_{\bar \alpha,0} ) = \partial_{\alpha} V_0|_{T=T_0} = 0\,.
\end{align} 
Plugging the results Eqs.~\eqref{eqs1} back into $V$ and introducing $\rho_{\alpha} = (T_{\alpha},
{\overline T}_{\bar \alpha})$ leads to the expression given in Eq.~\eqref{sp10}. From this, by minimizing we find the moduli displacements Eq.~\eqref{sp11}, and subsequently the most general expression for the effective inflaton potential, cf.~Eq.~\eqref{eq:GeneralEffPot},
\begin{align}\begin{split}
V &= \frac12 V_1 \left(T_{\alpha,0}, {\overline T}_{\bar \alpha,0}\right) m \varphi^2 
+ \frac14 V_2 \left(T_{\alpha,0}, {\overline T}_{\bar \alpha,0} \right) m^2 \varphi^4
\\ &- \frac12 
\begin{pmatrix} 
\frac{\partial V_1}{\partial T_{\alpha}} & \frac{\partial V_1}{\partial {\overline T}_{\bar \alpha}} 
\end{pmatrix}  
\begin{pmatrix} 
(m^{-2})^{\alpha {\bar \beta}} & (m^{-2})^{\alpha {\beta}}  \\ (m^{-2})^{\bar \alpha {\bar \beta}} & (m^{-2})^{\bar \alpha {\beta}}  
\end{pmatrix} 
\begin{pmatrix} 
\frac{\partial V_1}{\partial {\overline T}_{\bar \beta}} \\ \frac{\partial V_1}{\partial T_{\beta}}  
\end{pmatrix} 
m^2 \varphi^4 + \dots\,.
\end{split}\end{align}
By a straight-forward computation, one can find
\begin{align}\begin{split}
\frac{\partial V_1}{\partial T_{\alpha}}|_{T=T_0} = 
e^{K_0} &\bigg\{  - \frac{1}{2} K_0^{\beta \bar \gamma} 
\left[ K_{\bar \gamma} D_{\alpha} D_{\beta} W_\text{mod} + (K_{\alpha \beta} + K_{\alpha}
K_{\beta} - \Gamma_{\alpha \beta}^{\gamma} K_{\gamma}) \overline{D}_{\bar \gamma} \overline{W}_\text{mod} \right]  \\  
&+ D_{\alpha} W_\text{mod} + K_{\alpha} {\overline W}_\text{mod} + m K_1^{-2} (K_{\alpha} K_1 - K_{1, \alpha}) \bigg\} \,, \label{sp12} 
\end{split}\end{align}
where $D_{\alpha} D_{\beta} W = \nabla_{\alpha} D_{\beta} W + K_{\alpha} D_{\beta} W $. 

Using the mass formulas of Appendix~\ref{app:Masses}, the effective potential Eq.~\eqref{eq:GeneralEffPot} can be further simplified. In particular, using the approximation that the supersymmetric mass scale is much larger than $m_{3/2}$, cf.~Appendix \ref{app:Massapprox}, we find
\begin{align}
&\begin{pmatrix} 
\frac{\partial V_1}{\partial T_{\alpha}} & \frac{\partial V_1}{\partial {\overline T}_{\bar \alpha}} 
\end{pmatrix}  
\begin{pmatrix} 
(m^{-2})^{\alpha {\bar \beta}} & (m^{-2})^{\alpha {\beta}}  \\ (m^{-2})^{\bar \alpha {\bar \beta}} & (m^{-2})^{\bar \alpha {\beta}}  
\end{pmatrix} 
\begin{pmatrix} 
\frac{\partial V_1}{\partial {\overline T}_{\bar \beta}} \\ \frac{\partial V_1}{\partial T_{\beta}}  
\end{pmatrix} \nonumber \\
&\approx \frac{\partial V_1}{\partial T_{\alpha}} (m^{-2})^{\alpha {\bar \beta}} \left[  \frac{\partial V_1}{\partial {\overline T}_{\bar \beta}}  -m^2_{\bar \beta \bar \gamma} (m^{-2})^{{\bar \gamma} \beta} \frac{\partial V_1}{\partial T_{\beta}}  \right] + \text{h.c.}  \nonumber \\
&\approx \frac{1}{2} e^{K_0} K_0^{\alpha \bar \beta} K_{0, \alpha} K_{0, \bar \beta}   -\frac{1}{2} e^{3 K_0/2} \bigg\{   K_{\delta} (m_F^{-1})^{\beta \delta}
\Big[- K_0^{\epsilon \bar \epsilon} (K_{\beta \epsilon} + K_{\beta} K_{\epsilon}
- \Gamma_{\beta \epsilon}^{\gamma} K_{\gamma}) \overline D_{\bar \epsilon} \overline W_\text{mod} \nonumber \\
&\hspace{3.2cm}+ 2 D_{\beta} W_\text{mod} + 3 K_{\beta} {\overline W}_\text{mod}+2mK_1^{-2}\left(K_{0,\beta} K_1 - K_{1,\beta}  \right) \Big] + \text{h.c.}  \bigg\} \,.   \label{sp13} 
\end{align} 
Inserting this into $V$, we find the approximate effective potential Eq.~\eqref{sp14}. We remark that there are subtleties involved: when supersymmetry is broken, the fermion mass matrix has a zero eigenvalue, corresponding to the goldstino direction. Therefore, it is necessary to make the scalar partner of the goldstino very heavy, so that its entry in the inverse scalar mass matrix can be neglected and Eq.~\eqref{sp14} indeed can be used to obtain the leading-order result. However, it would be interesting to find an analogous expression to Eq.~\eqref{sp14} in the case that this is not possible. Note that this problem can be avoided in the case where the supersymmetry breaking field is nilpotent \cite{Antoniadis:2014oya}.

\subsection{Applying the general result: KKLT moduli stabilization}
\label{app:KKLTExample}

In order to illustrate how this general result can be applied to specific examples, we consider the effective action described by
\begin{align}\begin{split}
K &= - 3 \ln \left(T + {\overline T}\right) + \frac{1}{2} \left(\phi + {\bar \phi} \right)^2 + |X|^2 - \frac{|X|^4}{\Lambda^2} \,, \\
W &= W_0 + A e^{- a T} + f X + \frac12 m \phi^2 \,, \label{exp1}  
\end{split}\end{align}
i.e., the example of Sec.~\ref{sec:KKLT}. As discussed before, if $\Lambda$ is small enough the scalar in $X$ is heavy and its displacement during inflation $\delta X$ negligible. In this case, we can safely omit the sgoldstino and goldstino entries in the scalar and fermion mass matrices, respectively. With $f \sim \sqrt{3} W_0$ for cancellation of the cosmological constant, the leading-order vacuum expectation value of $T$ and its contribution to supersymmetry breaking are
\begin{align}
a A e^{-a T_0} + \frac{3}{2 T_0} W_0 \approx 0\, , \qquad G_T \approx - \frac{9}{4 a T_0^2}\,, \qquad G^T \approx - \frac{3}{a} \ . \label{exp2} 
\end{align} 
A leading-order computation of the scalar and fermion masses in Eqs.~\eqref{m1} and \eqref{m3}, respectively, leads to
\begin{align}
m_{T {\overline T}}^2 \approx 3 a^2 m_{3/2}^2 \,,\quad m_{TT}^2 \approx \frac{3 a}{2T_0} m_{3/2}^2 \,, \quad (m_F)_{TT} \approx - \frac{3 a}{2T_0} m_{3/2} \,. \label{exp3} 
\end{align}
Notice that $m_{TT}^2 \ll m_{T {\overline T}}^2$. Moreover, after canonical normalization of the kinetic terms the scalar and fermion masses are, at leading order 
\begin{align}
m_T^2 \approx |(m_F)_T|^2 \approx 4 a^2 T_0^2 m_{3/2}^2\,.
\end{align}
Therefore, in this case the ``nearly-supersymmetric" approximation outlined in Appendix~\ref{app:Massapprox} applies. Thus, also the final result Eq.~\eqref{sp14} does apply.  Actually, using the assumption that the scalar $X$ is heavy and that its vacuum expectation value is negligibly small, all indices in Eq.~\eqref{sp14} turn out to be modulus indices only. A straight-forward computation then yields
\begin{align}
V &= \frac{1}{16 T_0^3} \left[ \left( m^2 + 3 m W_0 - \frac{9 W_0}{a T_0} m \right) \varphi^2 - \frac{3}{8} \left(1 + \frac{6}{2 a T_0 }\right) m^2 \varphi^4  \right] \nonumber \\
 &= \frac{1}{2} \left( {\tilde m}^2 + 3 {\tilde m} m_{3/2} - \frac{9}{2 a T_0}
{\tilde m} m_{3/2} \right) \varphi^2 - \frac{3}{16} \left(1 + \frac{3}{a T_0}\right) {\tilde m}^2 \varphi^4  \,, \label{exp4} 
\end{align}
which, at leading-order, coincides with the naive expectation of Sec.~\ref{sec:General1}, and at next-to-leading-order with the result of Sec.~\ref{sec:AnalyticKKLT}.

\subsection{Chaotic inflation with a stabilizer field}
\label{app:Stabilizer}

The coefficients of the Taylor series in Eq.~\eqref{s7} are given by 
\begin{subequations}\begin{align}
V_0 &= e^{K}  \left\{  K_0^{\alpha \bar \beta}  D_{\alpha} W_\text{mod} \overline{D}_{\bar \beta} \overline{W}_\text{mod} + K_1^{-1} m^2 |S|^2 + \frac{|K_{S \bar S}W_\text{mod} S|^2}{K_{S \bar S} +K_{S \bar S S \bar S} |S|^2} - 3 |W_\text{mod}|^2 \right\} \,, \\
V_1 &= - \frac{1}{2} e^{K}  \bigg\{  K_0^{\alpha \bar \beta} \left( K_{0, {\bar \beta}} D_{\alpha} W_\text{mod} + K_{0, \alpha } \overline{D}_{\bar \beta} \overline {W}_\text{mod} \right) \nonumber \\ 
&\hspace{2cm} + \frac{K_{S \bar S} (W_\text{mod} + \overline{W}_\text{mod})(1+ K_{S \bar S}|S|^2)}{K_{S \bar S} +K_{S \bar S S \bar S} |S|^2} - 3 (W_\text{mod} + \overline{W}_\text{mod}) \bigg\} \,, \\
V_2  &= e^{K}  \left\{ \left(  K_0^{\alpha \bar \beta} K_{0, \alpha } K_{0, {\bar \beta} } - 3 \right) |S|^2 +\frac{1+ K_{S \bar S}|S|^2}{K_{S \bar S} +K_{S \bar S S \bar S} |S|^2}  \right\}  \,, \label{s5}
\end{align}\end{subequations}
where we have defined $D_{\alpha} W_\text{mod} = W_{\text{mod}, \alpha} + K_{0, \alpha } W_\text{mod}$. Expansion of these coefficients in $\delta T_{\alpha}$ and $\delta \psi$ leads to
\begin{subequations}\begin{align}
V_0 &= \frac{1}{2} 
\begin{pmatrix}
\delta T_{\alpha} & \delta {\overline T }_{\bar \alpha} 
\end{pmatrix}  
\begin{pmatrix}
m^2_{\alpha {\bar \beta}} & m^2_{\alpha \beta} \\
m^2_{{\bar \alpha} {\bar \beta}} & m^2_{{\bar \alpha} \beta} 
\end{pmatrix} 
\begin{pmatrix}
\delta {\overline T }_{\bar \beta} \\ \delta {T}_{\beta} 
\end{pmatrix} + m_S^2 \delta \psi^2 + \dots \,,  \\
V_1 &= V_1 (T_{0 \alpha}, {\overline T }_{0 \alpha} ) + \frac{\partial V_1}{\partial T_{\alpha}} \delta T_{\alpha} +  \frac{\partial V_1}{\partial {\overline T }_{\bar \alpha}} \delta {\overline T }_{\bar \alpha} + \dots \,, \\
V_2 (T_{\alpha}, {\overline T }_{\alpha} ) &= \frac{e^{K_0}}{K_{S \bar S}} + \delta \psi^2 e^{K_0}  \left( K_0^{\alpha \bar \beta} K_{0,\alpha }  K_{0,{\bar \beta} } - \frac{K_{S \bar S S \bar S} }{K_{S \bar S}^2 } \right) + \dots \,, \label{s6}
\end{align}\end{subequations}
keeping only the leading-order terms at order $\varphi^4$ in the scalar potential. Plugging the coefficients back into $V$ and minimizing with respect to the field displacements gives Eqs.~\eqref{s8} and \eqref{ss8}. The most general result for the inflaton scalar potential then reads
\begin{align}\label{sss8}\nonumber
V &= \frac12 m^2 \varphi^2\left[ \frac{e^{K_0} }{K_{S \bar S}} - \frac{V_1^2}{m_S^2 +  \frac12 m^2 \varphi^2 e^{K_0}  \left( K_0^{\alpha \bar \beta} K_{0 \alpha } K_{0 {\bar \beta} } - \frac{K_{S \bar S S \bar S} }{K_{S \bar S}^2 } \right)}  \right] \\
&\hspace{0.5cm} - \frac{m^4 \varphi^4 V_1^2}{2 \left[ m_S^2 +  \frac12 m^2 \varphi^2 e^{K_0}  \left( K_0^{\alpha \bar \beta} K_{0 \alpha } K_{0 {\bar \beta} } - \frac{K_{S \bar S S \bar S} }{K_{S \bar S}^2 } \right) \right]^2} \nonumber \\ 
&\hspace{1.2cm}\times 
\begin{pmatrix}
\frac{\partial V_1}{\partial T_{\alpha}} & \frac{\partial V_1}{\partial {\overline T }_{\bar \alpha}} 
\end{pmatrix}  
\begin{pmatrix}
(m^{-2})^{\alpha {\bar \beta}} & (m^{-2})^{\alpha {\beta}}  \\
m^{-2})^{\bar \alpha {\bar \beta}} & (m^{-2})^{\bar \alpha {\beta}} 
\end{pmatrix} 
\begin{pmatrix}
\frac{\partial V_1}{\partial {\overline T }_{\bar \beta}} \\ \frac{\partial V_1}{\partial T_{\beta}} 
\end{pmatrix} + \dots 
\end{align}
Again we can rewrite
\begin{align}\begin{split}
\frac{\partial V_1}{\partial T_{\alpha}}\bigg|_{T=T_0,S=0} &= - \frac{1}{2} e^{K_0} \left\{   K_0^{\beta \bar \gamma}  [ K_{\bar \gamma} D_{\alpha} D_{\beta} W_\text{mod} + (K_{\alpha \beta} + K_{\alpha} K_{\beta} - \Gamma_{\alpha \beta}^{\gamma} K_{\gamma}) \overline{D}_{\bar \gamma} \overline{W}_\text{mod} ]   \right. \\ 
& \hspace{2cm}\left. - D_{\alpha} W_\text{mod} - K_{\alpha} {\overline W}_\text{mod}  \right\} \,. \label{s9} 
\end{split}\end{align}
In the near-supersymmetric limit, using the expressions in Appendix~\ref{app:Massapprox}, we find
\begin{align}
&\begin{pmatrix}
\frac{\partial V_1}{\partial T_{\alpha}} & \frac{\partial V_1}{\partial {\overline T }_{\bar \alpha}} 
\end{pmatrix}  
\begin{pmatrix}
(m^{-2})^{\alpha {\bar \beta}} & (m^{-2})^{\alpha {\beta}}  \\
m^{-2})^{\bar \alpha {\bar \beta}} & (m^{-2})^{\bar \alpha {\beta}} 
\end{pmatrix} 
\begin{pmatrix}
\frac{\partial V_1}{\partial {\overline T }_{\bar \beta}} \\ \frac{\partial V_1}{\partial T_{\beta}} 
\end{pmatrix} \nonumber \\ \nonumber
& \approx \frac{1}{2} e^{K_0} K_0^{\alpha \bar \beta} K_{0, \alpha} K_{0, \bar \beta} +  \label{s10}  \frac{1}{2} e^{3 K_0/2} \bigg\{ K_{\delta} (m_F^{-1})^{\beta \delta} \Big[K_0^{\epsilon \bar \epsilon} (K_{\beta \epsilon} + K_{\beta} K_{\epsilon} - \Gamma_{\beta \epsilon}^{\gamma} K_{\gamma}) {\overline D}_{\bar \epsilon} {\overline W}_\text{mod}\\
&\hspace{5cm} -  D_{\beta} W_\text{mod} - \frac{1}{2} K_{\beta} {\overline W}_\text{mod} \Big] + \text{h.c.}  \bigg\} \,.  
\end{align} 
Using this, we find the simplified inflaton scalar potential in Eq.~\eqref{s11}.

\subsection{Details on K\"ahler Uplifting and LVS}
\label{app:KU}

From the K\"ahler potential of the K\"ahler Uplifting scenario, 
\begin{align}
K = -2 \ln\left[(T + \overline T)^{3/2} + \xi\right]\,, \nonumber
\end{align}
we obtain for its derivative and the inverse K\"ahler metric,
\begin{align}
K_T &= -\frac{3(T + \overline T)^{1/2}}{(T + \overline T)^{3/2} + \xi}\,, \label{KU1}\\
K^{T\overline T} &= \frac{(T + \overline T)^{1/2}}{3}\frac{((T + \overline T)^{3/2} + \xi)^2}{(T + \overline T)^{3/2} -\frac{\xi}{2}}\,. \label{KU2}
\end{align}
Note that $K^{T \overline T}$, and therefore the scalar potential, has a singularity at $T + \overline T = (\frac{\xi}{2})^{2/3}$.
To analyze the vacuum structure it is convenient to define the functions
\begin{align}
Y(T,\overline T) &= \frac{(-a + K_T)K^{T\overline T} K_T + 1}{\partial_T K^{T\overline T} + (-a + 2K_T) K^{T\overline T}} \,,\label{KU3}\\
Z(T,\overline T) &= {\frac{2}{Y(T,\overline T)^2}}\frac{K_T}{\partial_T K^{T\overline T} + (-a + 2K_T) K^{T\overline T}} \,. \label{KU4}
\end{align}
The Minkowski vacuum and the barrier to the run-away vacuum can be studied using an expansion in
$\eta= \xi/(2 (T + \overline T)^{3/2})$,
\begin{align}
K_T &= -\frac{3}{(T + \overline T)}(1 - 2\eta +\ldots)\,, \label{KU5}\\
K^{T\overline T} &= \frac{(T + \overline T)^2}{3}(1 + 5\eta +\ldots)\,, \label{KU6}\\
Y &= -\frac{3}{(T + \overline T)}\left[1 - {2\eta\left(1 + \frac{9}{4(a(T + \overline T) +4)}\right)} + \ldots\right]\,, \label{KU7}\\
Z &= \frac{2}{a(T + \overline T) + 4}\left[1 - 3\eta\left(1-\frac{9}{2(a(T + \overline T) +4)}\right) + \ldots\right]\,. \label{KU8}
\end{align}
The AdS minimum and the associated local maximum lie to the left of the singularity\footnote{Keeping in mind that this regime is not trustworthy from the perspective of supergravity and string theory.} where we can use an expansion for small $T + \overline T$,
\begin{align}
K_T &= - \frac{3(2T)^{1/2}}{\xi} \left[1 - \frac{(2T)^{3/2}}{\xi}+ \ldots\right] \,,\label{KU9}\\
{Y} &= - \frac{3(2T)^{1/2}}{2\xi}\left[1 {-\frac{10 (2T)^{3/2}}{\xi }}+ \ldots\right] \label{KU10} \ .
\end{align}

The simplest Large Volume Scenario is closely related to K\"ahler Uplifting. From the K\"ahler potential
\begin{align}
K = -2 \ln (\mathcal V + \xi) \,, \quad \mathcal V = (T_b + \overline T_b)^{3/2} - (T_\text s + \overline T_\text s)^{3/2}\,, \nonumber
\end{align}
corresponding to compactification on a swiss-cheese manifold, we obtain,
\begin{align}
K_\text b &= -\frac{3(T_\text b +\overline T_\text b)^{1/2}}{\mathcal V + \xi}\,, \label{LVS1}\\
K_\text s &= \frac{3(T_\text s +\overline T_\text s)^{1/2}}{\mathcal V + \xi}\,, \label{LVS2}\\
K^{\text b\bar {\text b}}  &= \frac{(T_\text b +\overline T_\text b)^{1/2}}{3}\frac{\left(\mathcal V + \xi\right) 
\left(\mathcal V + 3(T_\text s + \overline T_\text s)^{3/2} + \xi\right)}{\mathcal V -\frac{\xi}{2}}\,,\label{LVS3}\\
K^{\text b\bar {\text s}}  &= (T_\text b +\overline T_\text b) (T_\text s + \overline T_\text s)\frac{\mathcal V + \xi}{\mathcal V -\frac{\xi}{2}}\,,\label{LVS4}\\
K^{\text s\bar {\text s}}  &={\frac{1}{3}}(T_\text s +\overline T_\text s)^{1/2}\frac{\left(\mathcal V + \xi\right) 
\left({2}\mathcal V + {3}(T_\text s + \overline T_\text s)^{3/2} - \xi\right)}{\mathcal V - \frac{\xi}{2}}\,,\label{LVS5}
\end{align}
with $\partial_{T_\text s} \equiv \partial_\text s$,  $\partial_{T_\text b} \equiv \partial_\text b$. Notice the partial no-scale cancellation
\begin{align}
K^{\text b\bar {\text b}}K_\text b^2 + 2 K^{\text b\bar {\text s}}K_\text b K_\text s + K^{\text s\bar {\text s}}K_\text s^2 = 3 + \frac32 \frac{\xi}{\mathcal V - \frac{\xi}{2}} \,. \label{LVS6}
\end{align}
Since the superpotential does not depend on $T_\text b$ the scalar potential takes the simple form
\begin{align}
V = e^K\left[\left(K^{\text b\bar {\text b}}K_\text b^2 - 3\right)|W|^2 + K^{\text b\bar {\text s}}K_\text b \left(W\overline{D_{T_\text s}W} + D_{T_\text s}W\overline{W}\right) + K^{\text s\bar {\text s}}|D_{T_\text s}W|^2\right] \,.
\end{align}
Analogous to K\"ahler Uplifting the two equations for local extrema, $\partial_\text b V = \partial_\text s V = 0$, lead to two quadratic equations for $D_{T_\text s}W$,
\begin{align} 
A_i W^2 + B_i W D_{T_\text s} W + C_i (D_{T_\text s}W)^2 = 0 \,, \qquad i=1,2\,, \label{LVS7}
\end{align}
where we have assumed real parameters. If $W^2 \neq 0$ these can be rewritten as
\begin{align}
D_{T_\text s} W = \tilde Y W \,, \quad \tilde Z_i = 0\,, \nonumber
\end{align}
where 
\begin{align}
\tilde Y = \frac{A_1 C_2 - A_2 C_1}{B_2 C_1 - B_1 C_2}\,,\quad \tilde Z_i = A_i + B_i \tilde Y  + C_i \tilde Y^2 \,,\label{LVS8}
\end{align}
and
\begin{align}
A_1 &= K_\text b (K^{\text b\bar {\text b}} K_\text b^2 + 2K^{\text b\bar {\text s}} K_{b\bar s} - 3) + \partial_\text b (K^{\text b\bar {\text b}} K_\text b^2) \,,\label{LVS9}\\
B_1 &= 2(K^{\text b\bar {\text s}} K_\text b^2 + K^{\text s\bar {\text s}}K_{b\bar s} + \partial_\text b(K^{\text b\bar {\text s}}K_\text b))\,,\label{LVS10}\\
C_1 &= K_\text b K^{\text s\bar {\text s}} + \partial_\text b K^{\text s\bar {\text s}}\,,\label{LVS11}\\
A_2 &= \partial_\text s(K^{\text b\bar {\text b}}K_\text b^2) + K^{\text b\bar {\text s}}K_\text b (aK_\text s -{K_\text s^2}+ 2 K_{\text s\bar{\text s}})\,,\label{LVS12}\\
B_2 &= 2(K^{\text b\bar {\text s}}K_\text bK_\text s + \partial_\text s(K^{\text b\bar {\text s}}K_\text b)) + K^{\text b\bar {\text b}}K_\text b^2 - 3 \nonumber\\
&\quad- a K^{\text b\bar {\text s}} K_\text b + K^{\text s\bar {\text s}}(aK_\text s-{K_\text s^2} + 2K_{\text{s}\bar{\text s}})\,,\label{LVS13}\\
C_2 &= K^{\text s\bar {\text s}}({2}K_\text s - a) + K^{\text b\bar {\text s}}K_\text b + \partial_\text s K^{\text s\bar {\text s}}\,. \label{LVS14}
\end{align}
In the large volume expansion we obtain, with $T_\text b = \overline T_\text b$, $T_\text s = \overline T_\text s$,
\begin{align}
K_\text s &= \frac{3(2T_\text s)^{1/2}}{\mathcal V}\left[1 -{\frac{\xi}{\mathcal V}} + \ldots\right] \,,\label{LVS15}\\
\tilde Y &= {\frac{3(2T_\text s)^{1/2}}{\mathcal V}\left[1 + \frac{\xi}{2(2T_\text s)^{3/2}}-  {\frac{\xi}{4a(2T_\text s)^{5/2}}}+\dots \right]} \,,\label{LVS16}\\
\tilde Z_1 &= {\frac{9\xi}{4\mathcal V^{5/3}}} \left[1 - \frac{\xi}{(2T_\text s)^{3/2}} {-\frac{1}{aT_\text s}+\frac{\xi}{a(2T_\text s)^{5/2}}}+ \dots\right] \,,\label{LVS17}\\
\tilde Z_2 &= {\frac{3a\xi}{2\mathcal V}} \left[1 - \frac{\xi}{(2T_\text s)^{3/2}} {-\frac{5}{4aT_\text s}+\frac{3\xi}{2a(2T_\text s)^{5/2}}}+ \dots\right] \,.\label{LVS18}
\end{align}

\end{appendix}

%


\begin{thebibliography}{99}

\bibitem{Linde:1983gd} 
  A.~D.~Linde,
  Phys.\ Lett.\ B {\bf 129}, 177 (1983).


\bibitem{Ade:2013uln} 
  P.~A.~R.~Ade {\it et al.}  [Planck Collaboration],
  Astron.\ Astrophys.\  {\bf 571}, A22 (2014)
  [arXiv:1303.5082 [astro-ph.CO]].


\bibitem{Ade:2015tva} 
  P.~A.~R.~Ade {\it et al.}  [BICEP2 and Planck Collaborations],
  [arXiv:1502.00612 [astro-ph.CO]].


\bibitem{Kawasaki:2000yn} 
  M.~Kawasaki, M.~Yamaguchi and T.~Yanagida,
  Phys.\ Rev.\ Lett.\  {\bf 85}, 3572 (2000)
  [hep-ph/0004243].


\bibitem{Palti:2014kza} 
  E.~Palti and T.~Weigand,
  JHEP {\bf 1404}, 155 (2014)
  [arXiv:1403.7507 [hep-th]].


\bibitem{Marchesano:2014mla} 
  F.~Marchesano, G.~Shiu and A.~M.~Uranga,
  JHEP {\bf 1409}, 184 (2014)
  [arXiv:1404.3040 [hep-th]].


\bibitem{Hebecker:2014eua} 
  A.~Hebecker, S.~C.~Kraus and L.~T.~Witkowski,
  Phys.\ Lett.\ B {\bf 737}, 16 (2014)
  [arXiv:1404.3711 [hep-th]].


\bibitem{Grimm:2014vva} 
  T.~W.~Grimm,
  Phys.\ Lett.\ B {\bf 739}, 201 (2014)
  [arXiv:1404.4268 [hep-th]].


\bibitem{Ibanez:2014kia} 
  L.~E.~Ibanez and I.~Valenzuela,
  Phys.\ Lett.\ B {\bf 736}, 226 (2014)
  [arXiv:1404.5235 [hep-th]].


\bibitem{Blumenhagen:2014nba} 
  R.~Blumenhagen, D.~Herschmann and E.~Plauschinn,
  JHEP {\bf 1501}, 007 (2015)
  [arXiv:1409.7075 [hep-th]].


\bibitem{Hebecker:2014kva} 
  A.~Hebecker, P.~Mangat, F.~Rompineve and L.~T.~Witkowski,
  arXiv:1411.2032 [hep-th].


\bibitem{Ibanez:2014swa} 
  L.~E.~Ibanez, F.~Marchesano and I.~Valenzuela,
  JHEP {\bf 1501}, 128 (2015)
  [arXiv:1411.5380 [hep-th]].


\bibitem{Garcia-Etxebarria:2014wla} 
  I.~Garc'a-Etxebarria, T.~W.~Grimm and I.~Valenzuela,
  arXiv:1412.5537 [hep-th].


\bibitem{Covi:2008ea} 
  L.~Covi, M.~Gomez-Reino, C.~Gross, J.~Louis, G.~A.~Palma and C.~A.~Scrucca,
  JHEP {\bf 0806}, 057 (2008)
  [arXiv:0804.1073 [hep-th]].


\bibitem{Covi:2008cn} 
  L.~Covi, M.~Gomez-Reino, C.~Gross, J.~Louis, G.~A.~Palma and C.~A.~Scrucca,
  JHEP {\bf 0808}, 055 (2008)
  [arXiv:0805.3290 [hep-th]].


\bibitem{Dudas:2014pva} 
  E.~Dudas,
  JHEP {\bf 1412}, 014 (2014)
  [arXiv:1407.5688 [hep-th]].


\bibitem{Mazumdar:2014bna} 
  A.~Mazumdar, T.~Noumi and M.~Yamaguchi,
  Phys.\ Rev.\ D {\bf 90}, no. 4, 043519 (2014)
  [arXiv:1405.3959 [hep-th]].


\bibitem{Buchmuller:2014vda} 
  W.~Buchmuller, C.~Wieck and M.~W.~Winkler,
  Phys.\ Lett.\ B {\bf 736}, 237 (2014)
  [arXiv:1404.2275 [hep-th]].


\bibitem{Buchmuller:2014pla} 
  W.~Buchmuller, E.~Dudas, L.~Heurtier and C.~Wieck,
  JHEP {\bf 1409}, 053 (2014)
  [arXiv:1407.0253 [hep-th]].


\bibitem{Kallosh:2004yh} 
  R.~Kallosh and A.~D.~Linde,
  JHEP {\bf 0412}, 004 (2004)
  [hep-th/0411011].


\bibitem{Dudas:2012wi} 
  E.~Dudas, A.~Linde, Y.~Mambrini, A.~Mustafayev and K.~A.~Olive,
  Eur.\ Phys.\ J.\ C {\bf 73}, no. 1, 2268 (2013)
  [arXiv:1209.0499 [hep-ph]].


\bibitem{Wieck:2014xxa} 
  C.~Wieck and M.~W.~Winkler,
  Phys.\ Rev.\ D {\bf 90}, no. 10, 103507 (2014)
  [arXiv:1408.2826 [hep-th]].


\bibitem{Kachru:2003aw} 
  S.~Kachru, R.~Kallosh, A.~D.~Linde and S.~P.~Trivedi,
  Phys.\ Rev.\ D {\bf 68}, 046005 (2003)
  [hep-th/0301240].


\bibitem{Balasubramanian:2004uy} 
  V.~Balasubramanian and P.~Berglund,
  JHEP {\bf 0411}, 085 (2004)
  [hep-th/0408054].


\bibitem{Westphal:2006tn} 
  A.~Westphal,
  JHEP {\bf 0703}, 102 (2007)
  [hep-th/0611332].


\bibitem{Balasubramanian:2005zx} 
  V.~Balasubramanian, P.~Berglund, J.~P.~Conlon and F.~Quevedo,
  JHEP {\bf 0503}, 007 (2005)
  [hep-th/0502058].


\bibitem{Davis:2008fv} 
  S.~C.~Davis and M.~Postma,
  JCAP {\bf 0803}, 015 (2008)
  [arXiv:0801.4696 [hep-ph]].


\bibitem{Kallosh:2011qk} 
  R.~Kallosh, A.~Linde, K.~A.~Olive and T.~Rube,
  Phys.\ Rev.\ D {\bf 84}, 083519 (2011)
  [arXiv:1106.6025 [hep-th]].


\bibitem{Giddings:2001yu} 
  S.~B.~Giddings, S.~Kachru and J.~Polchinski,
  Phys.\ Rev.\ D {\bf 66}, 106006 (2002)
  [hep-th/0105097].


\bibitem{Kallosh:2014wsa} 
  R.~Kallosh and T.~Wrase,
  JHEP {\bf 1412}, 117 (2014)
  [arXiv:1411.1121 [hep-th]].


\bibitem{Lebedev:2006qq} 
  O.~Lebedev, H.~P.~Nilles and M.~Ratz,
  Phys.\ Lett.\ B {\bf 636}, 126 (2006)
  [hep-th/0603047].


\bibitem{O'Raifeartaigh:1975pr} 
  L.~O'Raifeartaigh,
  Nucl.\ Phys.\ B {\bf 96}, 331 (1975).


\bibitem{Cicoli:2013swa} 
  M.~Cicoli, J.~P.~Conlon, A.~Maharana and F.~Quevedo,
  JHEP {\bf 1401}, 027 (2014)
  [arXiv:1310.6694 [hep-th], arXiv:1310.6694].


\bibitem{Chialva:2014rla} 
  D.~Chialva and A.~Mazumdar,
  arXiv:1405.0513 [hep-th].


\bibitem{Talaganis:2014ida} 
  S.~Talaganis, T.~Biswas and A.~Mazumdar,
  arXiv:1412.3467 [hep-th].


\bibitem{Linde:2011nh} 
  A.~Linde, M.~Noorbala and A.~Westphal,
  JCAP {\bf 1103}, 013 (2011)
  [arXiv:1101.2652 [hep-th]].


\bibitem{Buchmuller:2014rfa} 
  W.~Buchmuller, V.~Domcke and K.~Schmitz,
  JCAP11(2014)006
  [arXiv:1406.6300 [hep-ph]].


\bibitem{Buchmuller:2014dda} 
  W.~Buchmuller and K.~Ishiwata,
  arXiv:1412.3764 [hep-ph].


\bibitem{Dong:2010in} 
  X.~Dong, B.~Horn, E.~Silverstein and A.~Westphal,
  Phys.\ Rev.\ D {\bf 84}, 026011 (2011)
  [arXiv:1011.4521 [hep-th]].


\bibitem{Kaloper:2011jz} 
  N.~Kaloper, A.~Lawrence and L.~Sorbo,
  JCAP {\bf 1103}, 023 (2011)
  [arXiv:1101.0026 [hep-th]].


\bibitem{Dubovsky:2011tu} 
  S.~Dubovsky, A.~Lawrence and M.~M.~Roberts,
  JHEP {\bf 1202}, 053 (2012)
  [arXiv:1105.3740 [hep-th]].


\bibitem{Kaloper:2014zba} 
  N.~Kaloper and A.~Lawrence,
  Phys.\ Rev.\ D {\bf 90}, no. 2, 023506 (2014)
  [arXiv:1404.2912 [hep-th]].


\bibitem{McAllister:2014mpa} 
  L.~McAllister, E.~Silverstein, A.~Westphal and T.~Wrase,
  JHEP {\bf 1409}, 123 (2014)
  [arXiv:1405.3652 [hep-th]].


\bibitem{Silverstein:2008sg} 
  E.~Silverstein and A.~Westphal,
  Phys.\ Rev.\ D {\bf 78}, 106003 (2008)
  [arXiv:0803.3085 [hep-th]].


\bibitem{McAllister:2008hb} 
  L.~McAllister, E.~Silverstein and A.~Westphal,
  Phys.\ Rev.\ D {\bf 82}, 046003 (2010)
  [arXiv:0808.0706 [hep-th]].


\bibitem{Dodelson:2013iba} 
  M.~Dodelson, X.~Dong, E.~Silverstein and G.~Torroba,
  JHEP {\bf 1412}, 050 (2014)
  [arXiv:1310.5297 [hep-th]].
 
  \bibitem{soni}
S.~K.~Soni and H.~A.~Weldon,
  Phys.\ Lett.\ B {\bf 126} (1983) 215;
 V.~S.~Kaplunovsky and J.~Louis,
  Phys.\ Lett.\ B {\bf 306} (1993) 269
  [hep-th/9303040];
 S.~Ferrara, C.~Kounnas and F.~Zwirner,
  Nucl.\ Phys.\ B {\bf 429} (1994) 589
   [Erratum-ibid.\ B {\bf 433} (1995) 255]
  [hep-th/9405188];
  E.~Dudas and S.~K.~Vempati,
  Nucl.\ Phys.\ B {\bf 727} (2005) 139
  [hep-th/0506172].

\bibitem{Antoniadis:2014oya}
  I.~Antoniadis, E.~Dudas, S.~Ferrara and A.~Sagnotti,
  Phys.\ Lett.\ B {\bf 733} (2014) 32
  [arXiv:1403.3269 [hep-th]];
 S.~Ferrara, R.~Kallosh and A.~Linde,
  JHEP {\bf 1410}, 143 (2014)
  [arXiv:1408.4096 [hep-th]];
 R.~Kallosh and A.~Linde,
  arXiv:1408.5950 [hep-th];
 G.~Dall'Agata and F.~Zwirner,
  arXiv:1411.2605 [hep-th].

\end{thebibliography}
\end{document}